\documentclass[11pt,a4paper]{article}
\usepackage{jheppub,pifont}

\newcommand{\tick}{{\color{blue}\ding{52}}}
\newcommand{\cross}{{\color{red}\ding{55}}}

\subheader{Submitted to Annual Review of Nuclear and Particle Science, Volume 63}

\title{Progress in the Determination of the\\Partonic Structure of the Proton}

\author[a]{Stefano Forte}
\author[b]{and Graeme Watt}
\affiliation[a]{Dipartimento di Fisica, Universit\`a di Milano and INFN, Sezione di Milano,\\ Via Celoria 16, I-20133 Milano, Italy}
\affiliation[b]{Institut f\"ur Theoretische Physik, Universit\"at Z\"urich,\\ Winterthurerstrasse 190, CH-8057 Z\"urich, Switzerland}
\emailAdd{forte@mi.infn.it}
\emailAdd{gwatt@physik.uzh.ch}

\abstract{
  We review the current state of the art in the determination of the
  parton substructure of the nucleon, as expressed in terms of parton
  distribution functions (PDFs), and probed in high-energy
  lepton--hadron and hadron--hadron collisions, and we assess
  their implications for current precision collider phenomenology,
  in particular at the Large Hadron Collider (LHC).
  We review the theoretical foundations of PDF determination: the way 
  cross sections are expressed in terms of PDFs using perturbative QCD
  factorization and evolution, the methodology used
  to extract PDFs from experimental data, and the way
  in which different physical
  processes can be used to constrain different PDFs.
  We summarize current knowledge of PDFs and the limitations in accuracy
  currently entailed for the computation of hadron collider processes, in
  particular at the LHC.
  We discuss the current main sources of
  theoretical and phenomenological uncertainties, and the direction of
  progress towards their reduction in the future.
}


\begin{document}

\begin{flushright}
  IFUM-1005-FT \\
  ZU-TH 02/13 \\
  21st January 2013 % (date of submission to journal)
\end{flushright}

\maketitle

%%%%%%%%%%

\section{PARTON DISTRIBUTIONS IN PERTURBATIVE QCD}
\label{sec:intro}

Parton distribution functions (PDFs) encode the information on the
substructure of hadrons, and in particular the nucleon, in terms of
partons. Partons are quarks and gluons,
the basic degrees of freedom of quantum chromodynamics (QCD),
the theory of strong interactions, 
as probed in ``hard'' processes, i.e.,~high-energy processes which
admit a perturbative description. The physical, measurable
cross section for any process with hadrons in the initial state is
determined by folding PDFs with the perturbatively
computable cross section that describes the interaction between
partons. It follows that  essentially no theoretical prediction at a
hadron collider is possible without PDFs:
indeed, the recent observation~\cite{Aad:2012gk,Chatrchyan:2012gu} of a
Higgs-like particle at the Large Hadron Collider (LHC)
crucially relied on the knowledge of PDFs.
The importance of an accurate knowledge
of PDFs has thus enormously increased with the advent of the LHC
as the frontier accelerator for the study of
fundamental interactions.  For example, PDF uncertainties are essential
for precision determination of Standard Model parameters, and in making
predictions for the production of possible new heavy particles in
theories beyond the Standard Model.

At the current stage of knowledge of strong interactions, 
parton distributions cannot be computed from first principles. They
are instead determined by comparing the PDF-dependent prediction for one or
more physical processes with its actual measured value: in this sense, their
determination is akin to the problem of the measurement of fundamental
constants, with the important complication that one has to determine a
set of functions, rather than a set of numbers.

The determination of PDFs has gone through various
stages which mirror the evolution of the theoretical and
phenomenological understanding of QCD. At a very early
stage~\cite{McElhaney:1973nj,Kawaguchi:1976wm,DeRujula:1976tz,
Johnson:1976xc,Gluck:1976iz,Hinchliffe:1977jy,Buras:1977yj}, 
parton distributions were determined through a combination of
model assumptions and the first experimental results on
deep-inelastic electron--nucleon scattering (DIS).
These determinations were semi-quantitative at best, and
they were aimed at showing the compatibility of the data with
the partonic interpretation of hard processes.

As the accuracy of the data and confidence in perturbative QCD
improved, a first determination of the gluon distribution
was achieved~\cite{Gluck:1980cp}: this is nontrivial, 
because the gluon does
not couple to leptonic probes, and it was thus determined indirectly,
from the scale dependence of deep-inelastic structure functions. Soon
thereafter, the  first PDF sets~\cite{Duke:1983gd,Eichten:1984eu} were
produced, based on consistent ``global fits'', i.e., the data--theory
comparison for a set of different lepton--hadron and hadron--hadron scattering
processes, chosen in order to maximize the information on PDFs. 
These analyses were all performed at leading order (LO), namely,
using the lowest perturbative order in QCD calculations, which was
accurate enough
for these sets to be widely used for phenomenology in the ensuing
decade, despite the early availability
of next-to-leading order (NLO) tools~\cite{Devoto:1983sh}. 

However, thanks to the availability of
 high-precision deep-inelastic scattering and hadron collider data, 
the use of NLO theory soon became
mandatory. Correspondingly, fairly
wide sets of data of a varied nature were increasingly used
as an input to the PDF
determination, in order to minimize
as much as possible the r\^ole of theoretical
prejudice~\cite{Martin:1987vw,dflm,Gluck:1989ze,Morfin:1990ck}. 
NLO parton sets evolved into standard analysis
tools and were constantly updated throughout the ensuing
decade.  In particular, the wealth of deep-inelastic data from the HERA
collider led to a considerable increase of both accuracy and kinematic
coverage, and  eventually led to global parton
sets (such as CTEQ5~\cite{Lai:1999wy} and MRST2001~\cite{Martin:2001es})
which could provide an input at an adequate level of accuracy to the
NLO QCD computations used both in tests of the
Standard Model and in searches for new physics.
These sets of PDFs, while differing in many technical details, shared
the basic underlying approach: a functional form for PDFs is assumed,
parametrized by a relatively small number of parameters, that are
determined by optimizing the fit of the computed observables to
the experimental data. The PDF
set of Reference~\cite{Gluck:1989ze} was produced along the same lines, but
introducing the extra
``dynamical'' assumption that at a sufficiently low scale parton
distributions become valence-like.

Once PDFs became a tool for precision physics, an estimate of the
uncertainty  on their knowledge became mandatory.  Previously, 
the only way of estimating the
uncertainty related to the parton distribution was to compare results
obtained with several parton sets: an especially unsatisfactory
procedure given that many possible sources of systematic bias are
likely to be common to several parton determinations. 
The first determinations of parton distributions with
uncertainties were obtained by only fitting to restricted data sets (typically
from a subset of deep-inelastic experiments), but retaining
all the information on the correlated uncertainties in the underlying data,
and propagating it through the fitting
procedure~\cite{dflm,Alekhin:1996za,Barone:1999yv,Botje:1999dj}. The 
need for a systematic approach to the determination of PDFs with
reliable uncertainties  was stressed in the seminal papers of
References~\cite{Giele:1998gw,Giele:2001mr}, where an entirely different
approach to parton determination was suggested, based on Bayesian inference.
This approach was never fully implemented, but the need for PDFs
with uncertainties based on global fits was generally recognized.

The problem was tackled in References~\cite{Pumplin:2002vw,Martin:2002aw},
where it was shown that in order to obtain statistically meaningful
results the conventional methodology had to be supplemented with an
unorthodox treatment of uncertainties, where the standard approach
must be supplemented by a suitable ``tolerance'' rescaling.
Once this is done, it is possible to determine ``error'' PDF sets
along with the central best-fit, 
which  allow for a determination of a one-sigma contour in parameter
space about the best fit.  PDFs with uncertainties have become
the standard ever since, and more refined versions of the
tolerance method have been used in subsequent global fits from the
MSTW~\cite{Martin:2009iq} and CTEQ~\cite{Lai:2010vv}  groups.

An alternative approach to PDF fitting was proposed in
Reference~\cite{Forte:2002fg}, and eventually led the NNPDF collaboration
to produce a first PDF set based
on DIS data~\cite{Ball:2008by} and then a PDF set from a global
fit~\cite{Ball:2010de}. This approach differs in two main respects
from the standard one.  The first is that PDFs are represented as a
Monte Carlo sample, from which the central value and uncertainty can
be computed respectively as a mean and standard deviation, rather than
from a best-fit and error sets. The second is that the functional form used
for the PDF parametrization, based on neural networks, has a very
large number of parameters (more than 250 for the PDF sets of
References~\cite{Ball:2008by, Ball:2010de}, to be compared to about 30 for
sets based on a standard parametrization). Therefore, the best-fit is
not determined as an absolute minimum of a figure of merit (such as
the $\chi^2$), which, given the large number of parameters, would
involve also fitting statistical noise, but rather by stopping the
minimization before the noise starts being fitted, through a suitable
criterion. 

PDF sets with uncertainties, based at least on NLO QCD
theory, and relying on a global set of data, had thus become the
standard by the late 2000s. However, the demands of precision
phenomenology, specifically at the LHC, have led to several
further theoretical and phenomenological improvements.
Firstly, with the increasing availability of calculations to
next-to-next-to-leading order (NNLO) in QCD, now all PDF sets
have been extended to include also sets which use
NNLO QCD theory in their determination. Furthermore, all
sets now include heavy-quark
mass effects. Finally, most sets are now available for a variety of values
of the strong coupling.

There are presently at least three sets of
PDFs with all these features 
which are being maintained and updated, from the CTEQ/CT,
MSTW and NNPDF collaborations. Further PDFs based on smaller data sets
have been produced recently: by the GJR/JR
group~\cite{Gluck:2007ck,JimenezDelgado:2008hf}, following the
``dynamical''  approach of Reference~\cite{Gluck:1989ze}; by the ABKM/ABM
group~\cite{Alekhin:2009ni,Alekhin:2012ig}, using mostly DIS data,
following the approach of
References~\cite{Alekhin:1996za,Alekhin:2002fv,Alekhin:2005gq}; and by the
HERAPDF group, which only uses HERA DIS
data~\cite{Aaron:2009aa,HERA:2010,HERA:2011}.

Several benchmarking exercises involving various sets of PDFs have
been performed
recently~\cite{Alekhin:2010dd,Alekhin:2011sk,
Watt:2011kp,Watt:2012np,Ball:2012wy},
and in particular the benchmarking of Reference~\cite{Alekhin:2011sk} has
led to the so-called PDF4LHC recommendation~\cite{Botje:2011sn} which
suggested the use of an envelope of the CTEQ/CT, MSTW and NNPDF PDFs for the
purposes of searches, calibration (e.g.,~acceptance computations) and 
precision tests of the Standard Model at the LHC, and in particular
for Higgs searches~\cite{Dittmaier:2011ti}.

The purpose of this review is twofold.  On the one hand it aims to provide an
accessible introduction to the theory and phenomenology of parton
distributions. 
In this respect the current review is more
concise and pedagogical than other recent
reviews~\cite{Forte:2010dt,DeRoeck:2011na,Perez:2012um} of the same or
related topics.
On the other hand, it aims to review the current state of the
art in PDF determination, to provide an assessment of their accuracy and
of the main sources of systematic and theoretical uncertainty on them,
and to discuss the impact they have on LHC phenomenology. In this
respect, it provides a more concise and critical snapshot than other
recent benchmark papers mentioned above.
This review updates the previous one,
over 20 years old and by now somewhat dated,
of Reference~\cite{Owens:1992hd}.

The structure of the remainder of the review is the following.
In Section~\ref{sec:det}
we will briefly review the theoretical framework which underlies the
definition of PDFs, the methodology used in their determination, and
the way in which individual data sets control different aspects of PDFs. In
Section~\ref{sec:status} we will summarize the current PDF fits,
their main features, and provide detailed comparisons between
them.  In Section~\ref{sec:pheno} we will discuss the impact
of PDFs, and particularly their uncertainties, on LHC phenomenology,
specifically by computing and comparing the computation of 
various precision LHC observables (``standard
candles'') with different PDF sets.  Finally, we comment on future
prospects in Section~\ref{sec:summary}.

%%%%%%%%%%

\section{PDF DETERMINATION}
\label{sec:det}

A PDF determination
involves first, obtaining a theoretical prediction for various
processes (at some given perturbative accuracy), and then comparing
this prediction to the data. The second step involves, in particular,
a methodology in order to extract PDFs and their uncertainties from
this comparison, and also, a choice of measurable processes in order to
maximize the information on the various PDFs.

\subsection{Theoretical Framework}
\label{sec:th}

The basic property which enables the perturbative computation of
cross sections for processes with hadrons in the initial state is
their factorization into a partonic cross section --- computed in
perturbation theory, using the quark and gluon degrees of freedom of
the QCD Lagrangian, and independent of the incoming hadron --- 
and parton distributions, which
characterize the hadronic bound states, and are universal, i.e., do not
depend on the specific process. 
Thanks to universality, it is possible to determine PDFs using the
experimental information on a particular set of processes, and then use them to
obtain predictions for different processes.  
Here we will review some basic
results, while referring to Reference~\cite{Ellis:1991qj} for a 
textbook treatment, and to Reference~\cite{Collins:2011zzd} for detailed
proofs of the underlying factorization theorems.

\subsubsection{Factorization for Hadroproduction}
\label{sec:hadrfact}

The cross section for a generic
hadroproduction process which depends on a single scale $M^2_X$
can be written in factorized form as
\begin{eqnarray}\label{hadrfact}
\sigma_X(s,M_X^2) &=& \sum_{a,b} \int_{x_{\rm min}}^1 {\rm d}x_1 \,
{\rm d}x_2 \, f_{a/h_1}(x_1,M_X^2) \,
f_{b/h_2}(x_2,M_X^2) \, {\hat \sigma_{ab\to X}}\left(
x_1x_2 s,M_X^2\right) \\
&=& \sum_{a,b} \sigma^0_{ab}  \int_{\tau}^1 \frac{{\rm d}x_1}{x_1}
 \int_{\tau/x_1}^1
\frac{{\rm d}x_2}{x_2} 
\, f_{a/h_1}(x_1,M_X^2) \,
f_{b/h_2}(x_2,M_X^2)\,C_{ab}\left(\frac{\tau}{x_1x_2},
\alpha_S(M^2_X)\right)\nonumber\\
&=& \sum_{a,b} \sigma^0_{ab}  \int_{\tau}^1\frac{{\rm d}x}{x}
\, {\cal L}_{ab}\left(x,M^2_X\right)\,
C_{ab}\left(\frac{\tau}{x},\alpha_S(M^2_X)\right),
\label{hadrfacta}
\end{eqnarray}
where $s$ is the center-of-mass energy of the hadronic collision, 
$f_{a/h_i}(x_i,M_X^2)$  is the
distribution of partons of type $a$ in the $i$th incoming hadron,
${\hat \sigma_{ab\to X}}$ is the parton-level cross section for
the production of the desired final state $X$,
the minimum value of $x_i$ is $x_{\rm min}= \tau$,
\begin{equation}
\tau\equiv\frac{M^2_X}{s}\label{taudef}
\end{equation}
is the scaling variable of the hadronic process, and in the last step leading to
Equation~\ref{hadrfacta} we defined the parton luminosity
\begin{equation}
{\cal L}_{ab}(x,M^2_X)\equiv \int_{x}^1 \frac{{\rm d}z}{z}
\,f_{a/h_1}\left(z,M_X^2\right)
\,f_{b/h_2}\left(\frac{x}{z},M_X^2\right)=\int_{x}^1
\frac{{\rm d}z}{z} \,f_{a/h_1}\left(\frac{x}{z},M_X^2\right)
\,f_{b/h_2}\left(z,M_X^2\right).
\label{lumdef}
\end{equation}
Equation~\ref{hadrfact} also  holds for factorizable multi-scale
processes (such as, say, Higgs production in $W$ fusion), with
$\sigma$ evaluated as a function of the incoming hadron momenta $p_1$
and $p_2$, and $\hat \sigma$ evaluated as a function of the incoming
parton momenta $x_1p_1$ and $x_2p_2$.

The hard coefficient function $C_{ab}\left(z,\alpha_S(M^2_X)\right)$ is 
 a function of the
scale $M_X^2$ and the dimensionless ratio  of this scale to the
center-of-mass energy $\hat s$ of the partonic subprocess:
\begin{equation}
z=\frac{M_X^2}{\hat s}=\frac{\tau}{x_1x_2},
\label{convvar}
\end{equation}
where $\tau$ is given by Equation~\ref{taudef}. A prefactor
$\sigma^0_{ab} $ has been extracted, so that at leading perturbative order the
coefficient function is either zero (for partons that do not couple to the
given final state at leading order), or else just a
Dirac delta:
\begin{equation} 
\hat \sigma_{ab\to X}=\sigma_0\,
C_{ab}\left(z,\alpha_S(M^2_X)\right),\qquad
C_{ab}\left(z,\alpha_S(M^2_X)\right)=c_{ab}\,\delta(1-z)+ \mathcal{O}(\alpha_S),
\label{cfdef}
\end{equation}
where the matrix $c_{ab}$ depends on the specific process.
For example, for virtual photon (Drell--Yan) production,
$c_{ab}$
is nonzero when $ab$ is a pair of a quark and an antiquark of the same
flavor, and in this case $\sigma_0=\frac{4}{9}\pi\alpha \frac{1}{s}$.
Equation~\ref{convvar} then implies that at leading order 
\begin{equation}\label{taulo}
\tau_{\rm LO}= x_1 x_2 .
\end{equation}

The factorized result of Equation~\ref{hadrfact}
generally holds both for inclusive
cross sections and rapidity
distributions. In the latter case, however, there is an extra
kinematic constraint which relates the hadronic and partonic kinematic
variables. In particular, at leading order the rapidity $Y$ of the
final state is related to the momentum fractions of the two partons by
\begin{equation}
Y_{\rm LO}=\frac{1}{2}\ln\frac{x_1}{x_2},
\label{lorapkin}
\end{equation}
which are thus completely determined by knowledge of $Y$ and $\tau$.

\subsubsection{Factorization for Electroproduction}
\label{sec:disfact}

For electroproduction, specifically deep-inelastic scattering, 
Equation~\ref{hadrfact} is replaced by a factorized expression for  
the structure functions $F_i(x,Q^2)$ that parametrize the inclusive
deep-inelastic scattering cross section:
\begin{equation}
\frac{{\rm d}^2\sigma^{{\rm NC},\ell^{\pm}}}{{\rm d}x\,{\rm d}Q^2}
(x,y,Q^2)=\frac{2\pi \alpha^2}{ x Q^4}
 \left[
Y_+ \,F_2^{\rm NC}(x,Q^2) \mp Y_- \,x F_3^{\rm NC}(x,Q^2)-y^2
\,F_L^{\rm NC}(x,Q^2)\right],
\label{eq:ncxsect}
\end{equation}
for neutral-current charged-lepton ($\ell^{\pm}$) DIS,  where 
the longitudinal structure function is defined as
\begin{equation}
F_L(x,Q^2)\equiv F_2(x,Q^2)- 2 xF_1(x,Q^2) ,
\label{eq:fdef}
\end{equation}
and
\begin{equation}
 Y_{\pm}\equiv 1\pm (1-y)^2,
\label{eq:ypmdef}
\end{equation}
in terms of the electron momentum fraction
\begin{equation}
y\equiv\frac{p\cdot q}{p\cdot k}=\frac{Q^2}{x s},
\label{ydef}
\end{equation}
and $p$ and $k$ are respectively the incoming proton and lepton
momenta, $q$ is the virtual photon momentum ($q^2=-Q^2$), and in the
last step, which holds neglecting the proton mass, $s$ is the
center-of-mass energy of the lepton--proton collision.
Similar expressions hold for charged-current scattering.

The factorized expression for the structure functions is
\begin{equation}
F_i(x,Q^2)=x\sum_a\int_x^1\frac{{\rm d}z}{z} \,C_{i,a}\left(\frac{x}{z},
\alpha_S(Q^2)\right)\,f_a(z,Q^2).
\label{disfac}
\end{equation}
Here, in the argument of the structure function
$x=\frac{Q^2}{2p\cdot q}$ is the standard Bjorken variable,
the hard coefficient function $C_{i,a}$ is the structure function computed
with an incoming parton, and $f_a(z,Q^2)$ is the distribution of the
parton $a$ in the only incoming hadron.
Also in this case at lowest
$\mathcal{O}(\alpha_S^0)$, the coefficient function $C_{i,a}$ is either zero
(for incoming gluons) or a constant (an electroweak charge)
times a Dirac delta.

\subsubsection{Perturbative Computations}
\label{sec:pert}

The factorized expressions in Equations~\ref{hadrfact} and \ref{disfac}
express the hadronic cross section in terms of PDFs at the same scale,
$M^2_X$ or $Q^2$, at which the hadronic cross section is
evaluated. However, PDFs at different scales are related by
perturbative evolution equations, namely the integro-differential equations
\begin{eqnarray}
&&\frac{\partial}{\partial\ln Q^2}\ \left (\begin{matrix} \Sigma(x,Q^2)
\\  g(x,Q^2)\end{matrix}\right)
=\int_x^1\frac{{\rm d}y}{y}
\left (\begin{matrix}P_{qq}^S\left(\frac{x}{y},\alpha_S(Q^2)\right) &
2n_fP_{qg}^S \left(\frac{x}{y},\alpha_S(Q^2)\right)\\
P_{gq}^S\left(\frac{x}{y},\alpha_S(Q^2)\right)
& P_{gg}^S\left(\frac{x}{y},\alpha_S(Q^2)\right) \end{matrix}\right)
\left (\begin{matrix}\Sigma(y,Q^2) \cr g(y,Q^2)\end{matrix}\right),\nonumber \\
&&\frac{\partial}{\partial\ln Q^2}\ q^{\rm NS}_{ij}\left(x,Q^2\right)
= \int_x^1\frac{{\rm d}y}{y} P^{\rm NS}_{ij}
\left(\frac{x}{y},\alpha_S(Q^2)\right) 
q^{\rm NS}_{ij}(y,Q^2),
\label{dglap}
\end{eqnarray}
where $g$ is the gluon distribution, $\Sigma$ denotes the singlet
quark distribution defined as
\begin{equation}
  \Sigma(x,Q^2)\equiv\sum_{i=1}^{n_f}
  \left(q_i(x,Q^2)+\bar q_i(x,Q^2)\right),
\label{singdef}
\end{equation}
and the nonsinglet quark distributions are defined as 
any linearly independent set of $2n_f-1$ 
differences of quark and antiquark distributions,
$q^{\rm NS}_{ij}(x,Q^2)=q_i(x,Q^2)-q_j(x,Q^2)$. The splitting functions
 $P_{ab}$ are perturbative series in $\alpha_S$,
that start at order $\alpha_S$ at LO.

There are some constraints on perturbative evolution due to
conservation laws, which hold at all scales: 
specifically the conservation
of baryon number
\begin{equation}
\int_0^1 \! {\rm d}x\left(q_i(x,Q^2)-
\bar{q}_i(x,Q^2)\right) = n_i\qquad(n_u=2, n_d=1, n_{s,c,b,t}=0),
\label{baryonsr}
\end{equation}
and the conservation of total energy-momentum
\begin{equation}
\int_0^1 \! {\rm d}x\,x \left[\sum_{i=1}^{n_f} \left(q_i(x,Q^2)+\bar
q_i(x,Q^2)\right)+ g (x,Q^2)\right]= 1.
\label{momsr}
\end{equation}

Combining the factorized expressions in
Equations~\ref{hadrfact} and \ref{disfac} 
with the solution to the evolution equations, 
physical observables can be  written as the convolution of a
prefactor, which contains both the coefficient functions and the
kernel that solves the evolution equations, with PDFs defined at some
reference scale. In all available determinations, PDFs are
parametrized at a fixed reference scale $Q_0^2$, and the solution to
the evolution
equations is used to produce tables of PDFs as a function of $x$
and $Q^2$ that are delivered to users, for example,
through the standard \textsc{lhapdf} interface~\cite{Whalley:2005nh}.

Several public codes for the solution of the evolution equations are
available~\cite{pegasus,hoppet,QCDnum}: these codes, as well as most of the
codes used internally by the various collaborations performing PDF
determinations, are benchmarked against standard
tables~\cite{lh2,heralhc}, originally produced by using two of
these codes~\cite{pegasus,hoppet} to evolve a set of reference toy PDFs.

Because PDFs are extracted from a particular set of processes,
and then used to make predictions for other processes, the
perturbative accuracy of the
predictions is limited by the perturbative accuracy of the computation
of the processes used in PDF determination. 
The accuracy of a perturbative QCD computation is fixed by the number
of orders which are included in the computation of the coefficient
functions in Equation~\ref{cfdef} (and their deep-inelastic
counterparts in Equation~\ref{disfac}), and of the splitting functions
$P_{ab}$ which
enter the evolution equations.  Leading order means
that both are computed to the lowest nonvanishing order, so splitting
functions
 to order $\alpha_S$ and coefficient functions to order
$\alpha_S^0$ for deep-inelastic scattering or Drell--Yan production, 
to order $\alpha_S^2$ for Higgs production via gluon fusion, and so on. 

Currently, splitting functions are known up to
NNLO~\cite{Moch:2004pa,Vogt:2004mw}, and coefficient functions are
known up to NNLO for several processes used for PDF determination, 
such as  Drell--Yan rapidity distributions~\cite{Anastasiou:2003yy},
though not yet for jet production (for DIS they are even known up to
N$^3$LO~\cite{Vermaseren:2005qc,Moch:2008fj}). PDFs may thus be
determined up to NNLO accuracy, although NLO PDFs are also important
because several collider processes are only known up to NLO. LO PDFs
can be useful for use in conjunction with Monte Carlo event
generators~\cite{Campbell:2006wx,Buckley:2011ms}, and are thus often
optimized for this
purpose~\cite{Sherstnev:2007nd,Lai:2009ne,Kasemets:2010sg}, for example,
by introducing some modifications of the standard LO expressions which
partly simulates the missing higher-order terms.

\subsubsection{Treatment of Heavy Quarks}

An important subtlety involves the treatment of heavy quarks. Indeed,
decoupling arguments~\cite{Appelquist:1974tg} imply that the
contribution of heavier
quark flavors to any process are power-suppressed at scales which are
below the threshold for their production~\cite{Collins:1978wz}. Therefore,
whereas in
principle the QCD Lagrangian contains six quark flavors, in practice
only a smaller number of ``active'' flavors are included in loops, and
thus in particular when determining the running of $\alpha_S$ and
solving the evolution equations.  When
expressing predictions for processes at various disparate scales in
terms of a single set of PDFs it is thus necessary to use a so-called
variable-flavor number (VFN) scheme, whereby different numbers of
active flavors are adopted at different scales.
Use of a fixed-flavor
number (FFN) scheme only allows comparison with the data in a
restricted range of scales.

In all PDF sets
currently in use, the PDFs for charm and heavier quarks are not
independently parametrized. Rather, heavy-quark distributions are
generated as the result of pair production from gluons, which, at
higher perturbative orders, can in turn be radiated from quarks. Such
terms may appear as contributions to the coefficient
functions, or as a result of perturbative evolution. If one varies the
scale $M^2_X$ or $Q^2$ at which PDFs are evaluated in the factorized 
expressions of Equations~\ref{hadrfact} or \ref{disfac}, they are
reshuffled between the coefficient function and the solution to
perturbative evolution equations. 

This entails a further
complication because, in the vicinity of the threshold for heavy-quark
production, the quark mass cannot be neglected. Whereas there is no
difficulty in principle in including the full dependence on the heavy-quark
masses in coefficient functions, the
solution of evolution equations only generates terms which depend
logarithmically on the heavy-quark mass itself. It is thus necessary to
explicitly include terms suppressed by powers of the heavy-quark mass  
in the coefficient functions, while subtracting the
logarithmically enhanced, unsuppressed terms that are already
generated by solving the evolution equations in order to avoid double
counting.

At present, there exist at least three different schemes to do so, all
of which have been worked out up to NNLO, i.e., including the exact
dependence on the heavy-quark mass up to $\mathcal{O}(\alpha_S^2)$, and all of
which are based on the renormalization scheme with explicit quark
decoupling of Reference~\cite{Collins:1978wz}:
ACOT~\cite{Aivazis:1993pi,Collins:1998rz}, 
recently extended to NNLO~\cite{Guzzi:2011ew},
TR~\cite{Thorne:1997ga,Thorne:1997uu,Thorne:2006qt},
and FONLL, originally proposed for
hadronic processes~\cite{Cacciari:1998it} and more recently extended
to DIS~\cite{Forte:2010ta}. These schemes have been benchmarked in
Reference~\cite{LHhq} (except ACOT at NNLO, then not available), where they
were shown to differ by subleading terms, which may not be entirely
negligible at NLO in the vicinity of the quark threshold, but rapidly
decrease at NNLO. Specifically, at NLO FONLL and ACOT coincide
exactly while TR differs by $\mathcal{O}(\alpha_S^2(m^2_c))$ terms;
at NNLO FONLL
and TR differ by $\mathcal{O}(\alpha_S^3(m^2_c))$ terms~\cite{LHhq}, while FONLL
and ACOT are expected~\cite{Guzzi:2011ew} to differ by
$\mathcal{O}(\alpha_S^3(Q^2))$.
There also exists a scheme (BMSN~\cite{Buza:1996wv}) which enables the
inclusion of a heavy flavor as active in the running of the coupling,
while not including it among the active flavors when solving evolution
equations.

\subsubsection{Electroweak Corrections}

So far we have only discussed higher order perturbative
corrections in the strong interactions. However, any process which
involves electroweak interactions, such as deep-inelastic scattering
or Drell--Yan production, also receives higher-order corrections in the
electromagnetic or weak interactions. These will enter in both
coefficient functions and evolution equations.
Roughly, because
at the electroweak scale the fine structure constant
$\alpha\sim\alpha^2_S\sim\frac{1}{100}$, one expects
NLO corrections in the electromagnetic interaction to become
relevant when computing at the NNLO order of the strong
interaction. Such corrections are not included in any of the most
recent PDF determinations, though in particular the effect of QED
corrections to evolution equations has been discussed in
References~\cite{Roth:2004ti,Martin:2004dh}. 

\subsection{Fitting Methodology}
\label{sec:fit}

Parton distributions are determined by comparing factorized
expressions of the form of Equations~\ref{hadrfact} and \ref{disfac} with
experimental data.  A confidence interval in the space of PDFs is then
determined by minimizing a suitable measure of goodness-of-fit. This
is nontrivial, because it involves defining a probability measure on a
space of functions~\cite{Giele:2001mr}. Two main methodologies to represent such
a probability measure are currently used: the Hessian approach and the
Monte Carlo approach. In any case, the infinite-dimensional problem of
representing a space of functions must be reduced to
a finite-dimensional form to be manageable, and this is done by
introducing a PDF parametrization, for which several choices are possible.

\subsubsection{Goodness-of-fit}
\label{sec:chisq}

Goodness-of-fit is measured by a $\chi^2$ function (see,
e.g., Reference~\cite{cowan}) 
\begin{equation} \label{eq:chisqcov}
  \chi^2 \;=\; \sum_{i=1}^{N_{\rm dat}}\sum_{i^\prime=1}^{N_{\rm dat}}
(D_i-T_i)\left(V^{-1}\right)_{ii^\prime}(D_{i^\prime}-T_{i^\prime}).
\end{equation}
Here, $D_{i}$ are the data points, $T_{i}$ are the theory
predictions, and 
the experimental covariance matrix
\begin{equation}\label{chi2mult0}
V_{ii^\prime} \;=\; \delta_{ii^\prime}\,(\sigma_{i}^{\rm uncorr})^2\;+\;
\sum_{k=1}^{N_{\rm corr}}\sigma_{k,i}^{\rm corr}\,
\sigma_{k,i^\prime}^{\rm corr},
\end{equation}
$i=1,\ldots,N_{\rm dat}$ labels the individual data points, generally
affected by  uncorrelated (statistical and systematic) uncertainties
$\sigma_{i}^{\rm uncorr}$, and $k=1,\ldots,N_{\rm corr}$ sources of
correlated systematic uncertainty $\sigma_{k,i}^{\rm corr}$.

Diagonal entries in the covariance matrix are simply the sum in
quadrature of all correlated and uncorrelated uncertainties: if
information on correlations is unavailable one may thus simply add
correlated and uncorrelated uncertainties in quadrature. However,
proper inclusion of correlations is necessary in order for the
$\chi^2$ to provide a faithful measure of goodness-of-fit,  as 
neglecting correlations
leads to an overestimation of uncertainties that may be substantial.
 A typical
situation where this may happen is when the correlated and
uncorrelated uncertainties are comparable in  size, and
$N_{\rm dat}\gg N_{\rm corr}$.

The $\chi^2$ of Equation~\ref{eq:chisqcov} can be rewritten by
introducing $N_{\rm corr}$ shift (or nuisance) parameters
$r_k$~\cite{Stump:2001gu,Pumplin:2002vw}:
\begin{equation} \label{eq:chisqcorr}
  \chi^2 \;=\; \sum_{i=1}^{N_{\rm dat}} \left(\frac{\hat{D}_{i}-T_{i}}
    {\sigma_{i}^{\rm uncorr}}\right)^2 \;+\; \sum_{k=1}^{N_{\rm corr}}r_{k}^2,
\end{equation}
where
\begin{equation} \label{eq:datashift}
  \hat{D}_{i} \equiv D_{i} - \sum_{k=1}^{N_{\rm corr}}r_{k}\,
  \sigma_{k,i}^{\rm corr} .
\end{equation}
Minimizing the $\chi^2$ in 
Equation~\ref{eq:chisqcorr} with respect to the shift parameters 
$r_{k}$ gives back Equation~\ref{eq:chisqcov}, so that these two
expressions are completely equivalent, and either can be used in practice.
The advantage of the expression in
Equation~\ref{eq:datashift} is that it is possible to study
the behavior of the shifts $r_k$ at the minimum: specifically, their 
distribution ought to be univariate Gaussian with mean zero.

There is a subtlety related to the possibility that some of the
uncertainties may be multiplicative, which becomes relevant if 
the best fit is determined by minimizing the $\chi^2$ of
Equation~\ref{eq:chisqcov}. An uncertainty is multiplicative if the
size of the uncertainty is proportional to the measured value, as
is the case, for example, for an overall normalization uncertainty. In
such case, it can be shown~\cite{D'Agostini:1993uj} that
minimization of the $\chi^2$ in
Equation~\ref{eq:chisqcov} would lead to biased results. Various ways
of dealing with this problem are discussed in Reference~\cite{Ball:2009qv},
and a recent summary of the approach adopted by various PDF fitting
groups is in the appendix of Reference~\cite{Ball:2012wy}.

\subsubsection{Parton Parametrization}

A set of PDFs is a set of functions, one for each parton entering the
factorized expressions in Equations~\ref{hadrfact} and \ref{disfac}. Because
PDFs at different scales are related by the evolution
equations, the goal is to determine a set of functions for
$0<x<1$ at some reference scale $Q_0^2$.

There are in principle thirteen
independent PDFs in a given hadron (six quarks and antiquarks and the
gluon); however, in practice, charm and heavier
quark PDFs in the nucleon are not independently determined
in all current PDF sets, and are instead
assumed only to be generated by QCD radiation. The (moderate) impact of 
introducing an independent (non-perturbative) charm PDF, so that
charm does not vanish below the threshold for its radiation (``intrinsic''
charm~\cite{Brodsky:1980pb}) has been studied in
References~\cite{Pumplin:2007wg,Martin:2009iq}.
While in the past some relations between
PDFs (such as, for example, equality of the strange and antistrange
PDFs) have been introduced by assumption, the standard for current
precision studies is to have a set of seven independent PDFs. In
practice, in many cases, it turns out to be convenient to express the
six light quark PDFs as suitable linear combinations,
like the singlet combination of Equation~\ref{singdef}.

Once a suitable set of basis PDFs has been chosen, all existing PDF
determinations are based on choosing a parametrization of PDFs at the
reference scale. A standard choice, adopted by most PDF fitting
groups, is to assume that 
\begin{equation}
f_i(x,Q_0^2)=x^{\alpha_i}(1-x)^{\beta_i} g_i(x),
\label{pdfparmgen}
\end{equation}
where $g_i(x)$ tends to a
constant for both $x\to 0$ and $x\to 1$. This choice is motivated by
the expectation that PDFs behave as a power of $x$ as $x\to0$ due to
Regge theory, and as a power of $(1-x)$ as $x\to1$ due to quark counting
rules (see, e.g., Reference~\cite{roberts} and references therein).
 Specific choices for the
function $g_i(x)$ differ between groups.  Common choices are a
polynomial or the exponential of a polynomial in $x$ or $\sqrt{x}$,
with  more parameters used to describe PDFs for which more information
is available (such as the gluon) in comparison to those (such as the
strange PDF) that are poorly constrained by the data.
Typical contemporary PDF sets based on this choice of functional form
are parametrized by about 20--30 parameters (see Section~\ref{sec:status} for a
detailed discussion). 

An altogether different option is to parametrize PDFs with a general
functional form which does not incorporate any theoretical
prejudice. Two options that have been considered recently are neural
networks~\cite{Forte:2002fg,Ball:2008by,Ball:2010de} and Chebyshev
polynomials~\cite{Glazov:2010bw}, though only
in the former case has a
full-fledged PDF set been constructed. In this context, neural
networks just provide a convenient unbiased set of (nonlinear) basis
functions. 
The neural networks used for PDF parametrization 
in References~\cite{Forte:2002fg,Ball:2008by,Ball:2010de} are 
multilayer feed-forward neural networks, one for each PDF and  all
with a fixed architecture (and thus number of parameters),
determined to be greatly redundant for the problem at hand.
In this case, the number of free
parameters is of order of 200--300. The intermediate option of
supplementing a parametrization of the form of
Equation~\ref{pdfparmgen} by a prefactor written as an expansion over
Chebyshev polynomials in order to study potential parametrization bias
has recently been explored in References~\cite{Pumplin:2009bb,Martin:2012da}.

When unbiased PDF parametrizations are adopted, and specifically when
the number of free parameters is very large, the absolute minimum of
the figure of merit is not necessarily the best fit, as this may
correspond to a result that might reproduce random fluctuations in
the data, or display oscillations which are unlikely to be present in
the (unknown) true result --- we will discuss briefly in
Section~\ref{sec:pdfunc} how this difficulty may be
circumvented. This clearly shows that the determination of a
set of functions from a finite set of data points is mathematically an
ill-posed problem, and thus that  the choice of a PDF parametrization is
a necessity. Whereas in all PDF determinations one tries to minimize
all sources of theoretical bias, a certain amount of theoretical
prejudice is thus always necessary in order to get a definite answer.

\subsubsection{Representation of PDF Uncertainties}
\label{sec:pdfunc}

There exist (at least) two commonly used ways of representing
probability distributions in the space of PDFs.
The first (so-called Hessian) option is based on the standard least-squares
method~\cite{cowan}. This is the procedure that is most commonly adopted when
using a  parametrization 
with a relatively small number of parameters. It is based on the
assumption that the probability distribution in the space of PDFs is
a multi-Gaussian in parameter space. Given a set of
experimental data points for a collection of processes that depend
on PDFs through factorized expressions of the form of
Equations~\ref{hadrfact} and \ref{disfac}, one first determines
a most likely PDF as the  best-fit PDF, in turn given 
by the set of parameters which minimizes
the $\chi^2$ in Equation~\ref{eq:chisqcov} for the data--theory
comparison (the way in which individual processes constrain specific PDFs is
discussed in Section~\ref{sec:data} below).

Once the best-fit has been determined, a confidence level (C.L.) about it is
determined by expanding the $\chi^2$ in parameter space about its minimum
to lowest nontrivial order.  The desired confidence level is
obtained as the volume in parameter space about
the minimum that corresponds to a fixed increase of the
$\chi^2$. For Gaussian uncertainties, the 68\% (or one-sigma) confidence level
corresponds to the volume enclosed by the $\chi^2=\chi^2_{\rm min}+1$
surface. This is called the Hessian method, because the
confidence level is entirely determined by the covariance matrix in
parameter space, which is the inverse of the (Hessian) matrix 
of second derivatives of the $\chi^2$ with respect to
the parameters, evaluated at the minimum.

In practice, in actual PDF fits involving large numbers of
experimental data points from different experiments, it turns out that the
textbook criterion of varying $\chi^2$ by $\Delta \chi^2=1 $  in order to
determine the one-sigma contour leads to unrealistic results. This
conclusion was arrived at~\cite{Collins:2001es,Pumplin:2002vw}
by comparing the parameter values that provide the best fit
to each set of experimental data: it is found
that these best-fit values fluctuate much more than one would expect
if $\Delta\chi^2=1$  did actually provide a 68\% confidence level in parameter
space. We will come back to the explanation for this fact, but we note
immediately that it could be due to
neglect or underestimate of one or more sources of data uncertainty. More
realistic results are obtained thus by assuming that the 68\%
confidence level is obtained by letting $\Delta \chi^2=T^2 $, where $T$
is a ``tolerance'' parameter, in turn determined by studying the
distribution of best-fit parameter values among experiments,
e.g.,~imposing that indeed 90\% of experiments approximately fall within
the 90\% confidence level. More refined methods involve determining a
different tolerance~\cite{Martin:2009iq} along each Hessian
eigenvector (``dynamical'' tolerance).

An obvious advantage of the Hessian method is that it allows for a
compact representation and computation of PDF uncertainties, by simply
providing eigenvectors of the Hessian matrix rescaled by their
respective eigenvalues, i.e., in practice, PDF sets which correspond
to the variation by a fixed amount (such as one-sigma, or 90\% C.L.) along the
direction of each eigenvector. PDF uncertainties on the PDFs
themselves, or any observables that depend on them, are then simply
found by adding in quadrature the variation along each direction.
So in a Hessian approach one delivers a central set of PDFs $S_0$, and
$N_{\rm par}$ one-sigma error sets $S_i$, corresponding to the variation
of each eigenvector in turn. The best-fit value of any quantity $F(S)$
which depends on the PDF set
(such as a cross section, or a PDF itself), and its
one-sigma uncertainty, are respectively:
\begin{equation} \label{hessonesig}
F_0=F(S_0), \quad 
\sigma_F=\sqrt{\sum_{i=1}^{N_{\rm par}} \left[F(S_i)-F(S_0)\right]^2}.
\end{equation}
(In practice, a slightly more complicated formula is often used that
gives asymmetric uncertainties).
The price to pay for this (besides the need to use linearized error
propagation) is that Hessian determination and diagonalization rapidly
become unmanageable if the number of parameters is too large.

An alternative way of representing  probability distributions in the
space of PDFs is the Monte Carlo method, whereby 
the probability
distribution of PDFs
is given by assigning a Monte Carlo sample of PDF replicas, namely
$N_{\rm rep}$ PDF sets $S^k$. 
Any feature  of the probability distribution can be determined from
the Monte Carlo sample. So, the best-fit value of any quantity $F(S)$
which depends on the PDF set
(such as a cross section, or a PDF itself) is now determined
as its expected value, namely  as the mean over the replica sample:
\begin{equation}\label{mccv}
F_0=\frac{1}{N_{\rm rep}}\sum_{k=1}^{N_{\rm rep}} F(S^k),
\end{equation}
while the one-sigma interval is now computed as a standard deviation
\begin{equation}\label{mconesig}
\sigma_F=\sqrt{\frac{1}{N_{\rm rep}-1}\sum_{k=1}^{N_{\rm rep}}
\left[F(S^k)-F_0\right]^2}.
\end{equation}
The obvious advantage of the Monte Carlo method is that it does not
require assumptions to be made on the form of the probability distribution in
parameter space, and also that
it provides a direct representation of the probability distribution,
which is convenient for many applications, as we shall see
shortly. 

There are various ways of constructing a Monte Carlo PDF replica sample.
One possibility~\cite{Forte:2002fg,Ball:2008by,Ball:2010de} is to
first construct a Monte Carlo representation of the starting data
sample. This means that, instead of giving a list of data points $D_i$ with the
covariance matrix $V_{ij}$ of Equation~\ref{chi2mult0}, one constructs a
  set of $N_{\rm rep}$  data replicas  $D_i^k$, with $i=1,\dots,N_{\rm dat}$  and
$k=1,\dots,N_{\rm rep}$, that reproduce the probability distribution of
  the data, i.e., such that the data points $D_i$ and
the generic element of the covariance matrix
can be respectively found by computing the average or the covariance
  over the replica sample:
\begin{eqnarray}\label{datmccv}
\langle D_i\rangle&\equiv& \frac{1}{N_{\rm rep}}\sum_{k=1}^{N_{\rm rep}}
D_i^k,\\
\label{datmccov}
{\rm cov}_{ij}&\equiv&\frac{1}{N_{\rm rep}-1}\sum_{i=1}^{N_{\rm rep}}
\sum_{j=1}^{N_{\rm rep}}\left(D_i^k -\langle D_i\rangle\right)
\left(D_j^k -\langle D_j\rangle\right).
\end{eqnarray}
One may verify a posteriori that when $N_{\rm rep}$ is large enough, then 
$\langle D_i\rangle$ tends to the experimental data points $D_i$,
and ${\rm cov}_{ij}$ tends to the experimental
covariance matrix $V_{ij}$ of Equation~\ref{chi2mult0}.
The Monte Carlo sample of PDFs is then determined by fitting a PDF set
$S^k$ to each data replica, which can be done as above by  minimizing a
suitable figure of merit. The set of data replicas is thus mapped onto
a set of PDF replicas.

This procedure is especially advantageous if PDFs are parametrized with a very
large number of parameters so that a reliable determination of the Hessian
matrix is impractical or impossible, because it only requires 
the determination of a best-fit PDF set for each data replica, without
full knowledge of the Hessian. 

However, even the determination of this
best-fit may be nontrivial if the number of parameters is very large: in this 
case, as mentioned, false minima and spurious fluctuations may
arise. Two methods to avoid this have been considered in the
literature. One is the cross-validation method~\cite{DelDebbio:2007ee}:
the data are randomly divided into two sets (``training'' and ``validation'');
the $\chi^2$ is then computed for both sets
separately, but only the $\chi^2$ of the training set is
minimized. Initially both the training and validation $\chi^2$
decrease, but at some point the training $\chi^2$ keeps decreasing
while the $\chi^2$ of the data in the validation
set starts increasing. The point at which this happens defines the best
fit. Picking a different partition of the data into training and
validation sets for each replica ensures that there is no information
loss, though of course this is only true in the limit of a large number
of replicas. 
The other method consists of adding to the $\chi^2$ a penalty term
that disfavors functional forms which are too complex: an option
that has been considered in the literature~\cite{Glazov:2010bw} is to
penalize PDFs which are longer with respect to a suitable metric (and
thus fluctuate too much). This procedure is perhaps more efficient,
but it  entails some subjectivity in  the choice of metric. At
present, only the cross-validation method has been implemented in a
full-fledged PDF determination~\cite{Ball:2008by,Ball:2011mu,Ball:2011uy}

Alternatively, one may view the Monte Carlo method as a
different way of delivering results that have already been obtained
by means of the Hessian method. In this case, one may
construct~\cite{Watt:2012tq} the PDF replicas $S^k$ by generating a
multi-Gaussian distribution of parameter values, centered at the best
fit and with width provided by the Hessian matrix itself, which is
easily done by choosing a basis in which the Hessian matrix is
diagonal, i.e.,
\begin{equation} \label{eq:randpdf}
F(S^k) = F(S_0) + \sum_{j=1}^{N_{\rm par}}\left[F(S_j)-F(S_0)\right]R_j^k,
\end{equation}
where $R_j^k$ is a random number taken from a univariate
Gaussian distribution with mean zero, and $S_0$ and $S_j$ are the usual
best-fit and eigenvector PDF sets.  In this case, it is possible
to verify a posteriori that the number of Monte Carlo PDF
replicas $N_{\rm rep}$ is large enough that the
original central value and Hessian covariance matrix are
reproduced. In practice it turns out that 
$N_{\rm rep}\sim50-100$ replicas are necessary and sufficient to either
reproduce the input data set of a typical present-day global
fit~\cite{Ball:2008by,Ball:2011mu,Ball:2011uy}, or its output Hessian
PDFs~\cite{Watt:2012tq}. 

There are several reasons why it is useful to construct a Monte Carlo
representation  of a PDF set, even if it has been determined in a
Hessian approach.
One reason is that once a Monte Carlo representation of a given PDF set is
available, new data can be included without performing a new fit, through
Bayesian reweighting, whereby the original Monte Carlo replicas are
supplemented by a weight which takes into account the effect of the
new data. The correct implementation of this technique, originally suggested in
Reference~\cite{Giele:1998gw}, was worked out in
References~\cite{Ball:2010gb,Ball:2011gg},
while in Reference~\cite{Watt:2012tq}
it was shown how it can be implemented in a Hessian fit: its only
limitation is that as new data are added, the number of starting
replicas should be increasingly large for the accuracy of the Monte
Carlo prediction to be preserved. On the other hand, it was pointed out in
Reference~\cite{Giele:2001mr} that if the number of new data included in
this way is increasingly large, then the dependence on the original
PDF set and parametrization becomes increasingly weak, and thus all
issues related to the choice of PDF parametrization (such as potential
bias) become increasingly less relevant.

Another reason why a Monte Carlo representation is useful is that it
provides a simple way of combining results obtained by different
groups. If, in particular, different groups arrive at independent PDF
determinations using the same (or almost the same) data and theory,
with differences only being due to either methodological choices, or
theoretical differences which are beyond the accuracy of the
calculation (such as, for example, different NNLO terms in a NLO
computation) there is a priori no way of deciding which group provides
the most reliable determination. An effective way of combining
results, while keeping into account the possibility of methodological
differences, and thus arriving at a more reliable result, is to
simply produce a Monte Carlo set in which an equal fraction of
replicas comes from each of the various
groups~\cite{Watt:2012tq,Forte:2010dt}, as we shall see more explicitly
in Section~\ref{sec:combination} below.  This statistical combination
is meaningful even, or especially, if the PDF sets from the different groups
are strongly correlated.  Moreover, the fraction taken from each group
need not be equal if it is desired to include the PDF set from one or more
groups with a different weight from the others.

\subsubsection{PDF Uncertainties: the State of the Art}

The availability of several distinct methodologies for PDF
parametrization and determination allows for a comparison of
results. Detailed comparisons will be presented in
Section~\ref{sec:status} below, but we note immediately that results
obtained with parametrizations of the form of Equation~\ref{pdfparmgen} and a
Hessian method with a tolerance criterion to determine uncertainty
bands are generally in reasonably
good agreement with those obtained using very general parametrizations
and a Monte Carlo method with a stopping criterion to determine
replica best fits. Also, both seem in good agreement with results
obtained using Bayesian reweighting, which do not depend on a fitting 
procedure. This provides good evidence that current estimates of 
PDF uncertainties are not too far off the mark. 

However, it would be interesting to understand
in more detail how the statistical features of the underlying data
propagate onto the statistical features of PDF sets. In particular, it
would be interesting to understand the detailed reasons for the need
to introduce tolerance in the Hessian procedure, and how it relates to
the distribution of  best-fit replicas when the  Monte Carlo method is used
in conjunction with a very general PDF parametrization.  Note that,
in principle, tolerance could be introduced also in the Monte Carlo approach
simply by rescaling experimental uncertainties during the generation of
data replicas.

It is clear that (at least) two different reasons may explain the need for
tolerance. One is the presence of data inconsistencies, or equivalently,
neglected or underestimated sources of uncertainty in PDF fits based
on very broad data sets. The second is the fact that the choice of PDF
parametrization is restricting the space of accessible PDFs. Various
investigations of the relative importance of these effects have been
presented: based on the Hessian approach with a standard~\cite{Pumplin:2009sc}
or extended Chebyshev~\cite{Pumplin:2009bb,Martin:2012da} parametrization,
or on the Monte Carlo approach with a standard~\cite{Watt:2012tq} or
neural network~\cite{Ball:2011eq} parametrization.
These studies shed light on the relevant
issues, but no consensus has yet
emerged on the relative impact of these effects on tolerance and on the
precise relation between tolerance and the way in which cross-validated fits
explore the space of PDF minima.
While the current accuracy in the determination of PDF uncertainties
is most likely acceptable if compared to other sources of uncertainty
in the computation of collider processes, the needs for greater
accuracy which come from the availability of higher-order
computations, and the hope of discovering new physics effects in small
deviations between the data and current predictions, will require
a deeper level of understanding of these issues.

PDF uncertainties, as we have discussed them so far, are those that follow
from propagation of the uncertainty of the experimental data that
underlie the PDF determination. As the accuracy increases, however,
other sources of uncertainties, and in particular all uncertainties
related to the theory used in PDF determination, become relevant. At
present, the only way of dealing with such uncertainties is to make
sure that they are small enough. For example, the impact of different
choices of treatment of heavy-quark masses was studied in
References~\cite{LHhq,Thorne:2012az}. The impact of higher order corrections
is studied by comparing NLO to NNLO PDFs. Higher-twist
(power-suppressed) corrections are kept under control by removing
data below some low cutoff scale that may be affected by them,
and their impact can be
studied by varying this cutoff~\cite{Martin:2003sk}. Nuclear corrections,
that affect some deep-inelastic scattering data in which targets are
deuterons or heavier nuclei, rather than just
protons, have been studied by
including such corrections according to various
models~\cite{Martin:2009iq,Ball:2009mk}, or by attempting to fit
the corrections directly~\cite{Martin:2009iq,Martin:2012da}.  In the future,
a more systematic approach to each of these sources of theoretical
uncertainty may be desirable, and in particular it may become
necessary to provide PDF sets
with an estimate of the theoretical PDF uncertainty.

\subsection{Data Constraints Before the LHC}
\label{sec:data}

We have seen that a typical PDF set includes  seven different
PDFs. This means that at least seven independent physical processes
for given kinematics are needed in order to determine all the PDFs.
In principle, this could be done by using deep-inelastic scattering
alone, though in practice it is convenient to use a broad combination
of data from both electro- and hadroproduction in order to obtain
accurate results. Here we will discuss how pre-LHC data can be used to
determine PDFs, while the impact of LHC data will be discussed in
Section~\ref{sec:pheno} below.

A primary r\^ole is played by  DIS and Drell--Yan data. To see this, 
note that the factorized expressions in Equations~\ref{hadrfact}
and~\ref{disfac} immediately imply that at leading order
deep-inelastic structure functions and Drell--Yan rapidity
distributions provide a direct handle on
individual quark and antiquark PDFs (DIS), or pairs of PDFs
(Drell--Yan). It is thus possible to understand what is dominantly 
measured by each individual process by looking at the leading order
 expressions.

The leading order contributions to the DIS  structure functions $F_1$ and
$F_3$ are  (at leading order $F_2=2 x F_1$):
\begin{equation}
\begin{tabular}[c]{lc}
NC\qquad&$F_1^{\gamma} =\sum_{i}  e^2_i\left(q_i+\bar
q_i\right)$\quad\qquad\\
NC\qquad&$F_1^{Z,\gamma Z} =\sum_{i}  B_i\left(q_i+\bar
q_i\right)$\quad\qquad\\
NC\qquad&$F_3^{Z,\gamma Z} =\sum_{i}  D_i\left(q_i+\bar
q_i\right)$\quad\qquad\\
CC\qquad&{ $F_1^{W^+} =\bar u + d + s + \bar c$}\quad\qquad\\
CC&${ -F_3^{W^+}/2 = \bar u - d - s +\bar c }$,\quad\qquad \\
\end{tabular}
\label{strfun}
\end{equation}
where NC and CC denotes neutral- or charged-current scattering and
the contributions coming from $Z$ exchange and
from $\gamma Z$ interference have couplings 
\begin{eqnarray}
B_q(M^2_X)&=& -2e_qV_\ell
V_qP_Z+(V_\ell^2+A_\ell^2)(V_q^2+A_q^2)P_Z^2,\nonumber \\
  D_q(M^2_X)&=&-2e_qA_\ell A_qP_Z+4V_\ell A_\ell V_qA_q P_Z^2,
\label{bdcoup}
\end{eqnarray}
where the electroweak couplings of quarks and leptons can be found,
e.g.,~in Reference~\cite{Forte:2010dt} and  $P_Z= M^2_X/(M^2_X+M_Z^2)$. 
The leading order contribution to Drell--Yan production is
\begin{equation}
\begin{tabular}[c]{lc}
$\gamma$\qquad & $\frac{{\rm d}\sigma}{{\rm d} M_X^2 {\rm d} y}(M_X^2,y) =
\frac{4\pi \alpha^2}{9 M_X^2 s}\sum_i e_i^2 L^{ii}(x_1,x_2)$\\
$W$\qquad& $\frac{{\rm d} \sigma}{{\rm d} y} = 
\frac{\pi G_F M_V^2\sqrt{2}}{3 s}
    \sum_{i,j} |V^{\rm CKM}_{ij}| L^{ij}(x_1,x_2)$\\
$Z$\qquad&$\frac{{\rm d} \sigma}{{\rm d} y} =  \frac{\pi G_F M_V^2\sqrt{2}}{3 s}
    \sum_i \left(V^2_{i}+A^2_i\right) L^{ii}(x_1,x_2)$
\end{tabular}
\label{lody}
\end{equation}
in terms of the differential leading order parton luminosity
\begin{equation}
L^{ij}\left(x_1,x_2\right)\equiv q_i(x_1,M_X^2)\, \bar
q_j(x_2,M_X^2)+q_i(x_2,M_X^2)\, \bar q_j(x_1,M_X^2)
\label{loplum}
\end{equation}
and the CKM matrix elements $V_{ij}$.

It follows that
a determination of deep-inelastic structure functions $F_1$ and
$F_3$  for charged-current deep-inelastic scattering provides four
independent linear combinations of quark
distributions (if $W^\pm$ can be distinguished),
with two more linear combinations provided by neutral-current
structure functions: all individual light quark and antiquark flavors
can then be determined by linear combination. This situation
would be realistic at a neutrino factory with both neutrino and
antineutrino beams and the possibility of identifying the charge of the
final state
lepton on an event-by-event basis~\cite{Mangano:2001mj}.
Unfortunately, this theoretically and phenomenologically very clean
option is at best far in the future, so at present the information on
individual PDFs can only be achieved by combining information from
different processes, each of which provides independent information, as
we shall now discuss.

\subsubsection{Isospin Singlet and Triplet}
\label{isospinsingt}

In neutral-current deep-inelastic scattering, only
the charge-conjugation even combination $q_i+\bar q_i$ can be determined.
 Specifically,
photon DIS data only determine the fixed combination in which each
flavor is weighted by the square of the electric charge, see
Equation~\ref{strfun}. However,
an independent combination may be accessed by also measuring DIS on a
neutron (in practice deuterium) target, and
 using isospin symmetry to relate the quark and antiquark
 distributions of the proton and neutron:
\begin{equation}
  { u}^{ p}(x,Q^2)= {d}^{ n}(x,Q^2),\quad
  { d}^{ p}(x,Q^2)= {u}^{ n}(x,Q^2).\label{isospin}
\end{equation}
One then has 
\begin{equation}
  { F_2^p(x,Q^2)-F_2^n(x,Q^2)= \frac{2}{3}\left[\left(u^p+\bar
        u^p\right)-\left(d^p+\bar d^p\right)\right]\left[1+
      \mathcal{O}(\alpha_S)\right]}\label{nonsingf2}
\end{equation}
so that the difference of proton and  neutron  structure functions
provides a leading-order handle on the isospin triplet combination
\begin{equation}
T_3(x,Q^2)\equiv u(x,Q^2)+\bar
    u(x,Q^2) -\left[d(x,Q^2)+\bar d(x,Q^2)\right].
\label{tripdef}
\end{equation}

\subsubsection{Light Quarks and Antiquarks}
\label{lightqqbar}

DIS data from HERA are available both for NC and CC scattering, both
with electron and positron beams. Unfortunately, collider data only
provide a fixed combination of the structure functions $F_1$ and $F_3$,
because for given $x$ and $Q^2$ Equation~\ref{eq:ncxsect} implies that
$y$ can be varied only by changing the center-of-mass energy of the
lepton--hadron collision.  Hence, HERA data only provide three
independent combinations of structure functions and thus of parton
distributions (NC and CC with positively or negatively charged
leptons). However, a fourth combination may be obtained  because the
$Q^2$ dependence of the $\gamma^*$ and $Z$ contributions to NC
scattering is different (see Equation~\ref{bdcoup}). It follows that the
very precise HERA data can determine four independent linear
combinations of PDFs, which can be chosen as the two lightest flavors
and antiflavors.

Currently, neutrino deep-inelastic
scattering data are available on heavy, approximately isoscalar,
nuclear targets. The energy of the neutrino beam usually has
a spectrum, so the value of $y$ given by Equation~\ref{ydef} is
not fixed, and the contributions of $F_1$ and $F_3$ to the cross
section can be disentangled. On an isoscalar target at leading order
\begin{eqnarray}
  F_2^{\nu}&=& x( u+\bar{u}+d+\bar{d}+2s
  +2\bar{c} )+\mathcal{O}(\alpha_S), \nonumber \\
  F_2^{\bar{\nu}} & =& x( u+\bar{u}+d+\bar{d}
  +2\bar{s}+2c)+\mathcal{O}(\alpha_S),\nonumber \\
  F_3^{\nu}&=&  u-\bar{u}+d-\bar{d}+2s-2\bar{c}+\mathcal{O}(\alpha_S),
  \nonumber \\
  F_3^{\bar{\nu}} &=& u-\bar{u}+ d-\bar{d}-2\bar{s}+2c+\mathcal{O}(\alpha_S),
  \label{f23nu}
\end{eqnarray}
so neutrino data provide an accurate handle on the total valence component
\begin{equation}
V(x,Q^2)=\sum_{i=1}^{n_f} (q_i(x,Q^2)-\bar q_i(x,Q^2)).
\label{valdef}
\end{equation}

A more direct determination of the light flavor decomposition can be
obtained using  the Drell--Yan process, and in particular by selecting
different PDF combinations, by looking
at different final states. Specifically~\cite{Ellis:1990ti},
for neutral-current Drell--Yan production on
proton and neutron (or deuteron) targets, using the isospin symmetry in
Equation~\ref{isospin} one gets at leading order 
\begin{equation}
\frac{\sigma^{pn}}{\sigma^{pp}}\sim \frac{\frac{4}{9}u^p\bar
  d^p+\frac{1}{9}d^p\bar u^p}{\frac{4}{9}u^p\bar
  u^p+\frac{1}{9}d^p\bar d^p}+ \mathcal{O}(\alpha_S)+\hbox{heavier quarks},
\label{dyncasym}
\end{equation}
where ``heavier quarks'' denotes strange and heavier flavors,
which give a smaller
contribution at least in the region of $x\gtrsim 0.1$.
In particular, in the ``valence'' region   $x\gtrsim 0.1$  the
up distribution is roughly twice as large as the down distribution
(assuming $\bar u\sim\bar d$), so the
first term in both the numerator and the denominator of
Equation~\ref{dyncasym} gives the dominant contribution, and the ratio
reduces to $ \frac{\sigma^{pn}}{\sigma^{pp}}\approx \frac{\bar
  d^p}{\bar u^p}$.
Hence this particular combination of cross sections provides a sensitive
probe of the $\bar u/\bar d$ ratio.

In the charged-current case, one may exploit the fact that
using charge-conjugation symmetry to relate the $p$ and $\bar p$ PDFs,
\begin{equation}
q_i^{p}=\bar q_i^{\bar p},
\label{ccon}
\end{equation}
at leading order one gets 
\begin{equation}
\frac{\sigma^{p\bar p}_{W^+}}{\sigma^{p\bar p}_{W-}}=\frac{ u^p(x_1)
  d^p(x_2)+ \bar d^p(x_1)\bar u^p(x_2)}{d^p(x_1)
  u^p(x_2)+ \bar u^p(x_1)\bar d^p(x_2)}+ \mathcal{O}(\alpha_S)+\hbox{ Cabibbo
  suppressed}+\hbox{ heavy quarks}
\label{dywasym}
\end{equation}
where heavy quarks denotes charm and heavier flavors, and
we have assumed that cross sections are
differential in rapidity. If the kinematics are chosen in such a 
way that $x_i$ are in the ``valence''
region, where quark distributions are sizably larger than antiquark
ones, then the ratio of Equation~\ref{dywasym} is mostly sensitive to 
the light quark ratio $u/d$~\cite{Berger:1988tu,Martin:1988aj}.

\subsubsection{Strangeness}
\label{strangesec}

Strangeness is nontrivial to determine,  because it has the same
electroweak couplings as the down distribution, while it is typically
smaller than it. The only way of determining it
accurately from DIS is to include
semi-inclusive information. A simple way of doing this is to use data
for neutrino deep-inelastic charm production (known as ``dimuon''
production). 
At leading order the structure functions are then
\begin{equation}
  F_2^{\nu p,c}(x,Q^2)=x F_3^{\nu p,c}(x,Q^2)= 2x\,
  \big(|V_{cd}|^2\,d(x)\,+|V_{cs}|^2\,s(x)+|V_{cb}|^2\,b(x)\big)+
  \mathcal{O}(\alpha_S),
  \label{fnbexpr}
\end{equation}
while $F_2^{\bar{\nu} p,c}(x,Q^2)=-x F_3^{\bar{\nu} p,c}(x,Q^2)$ probes the
corresponding antiquarks.

Drell--Yan data also constrain
strangeness. Specifically, the cross-section ratio of Equation~\ref{dywasym}
receives a contribution from strange and charm quarks which, up to CKM
matrix elements, coincides with the contribution from down and up
quarks respectively. Well above the charm threshold this contribution is
sizable, so comparing high- and low-scale Drell--Yan data 
potentially leads to a rather accurate determination of
strangeness~\cite{Ball:2010de}.

\subsubsection{Gluons}
\label{glusec}

The determination of the gluon distribution is nontrivial because the
gluon does not couple to electroweak final states. It does, however,
mix at leading order through perturbative evolution: so even in LO 
expressions for cross sections and structure functions, the
gluon does determine their scale dependence. Indeed
\begin{equation}
\frac{\partial}{\partial\ln Q^2} F_2^S (x,Q^2)=\int_x^1 \frac{{\rm d}y}{y}
 \left[{ P_{qq}^S\left(\frac{x}{y},\alpha_S(Q^2)\right)}
{ F_2^{S}(y,Q^2)}+2  \, 
n_f{P_{qg}^S\left(\frac{x}{y},\alpha_S(Q^2)\right)g(y,Q^2)}\right]+
\mathcal{O}(\alpha_S^2),
\label{f2loscaldep}
\end{equation}
where by $F^S_2(x,Q^2)$ we denote the singlet 
component (defined as in Equation~\ref{singdef}) of the $F_2$ structure
function.

It follows that the gluon is mostly determined by scaling
violations, or by its coupling to strongly-interacting final states,
i.e., jets. The main shortcoming of the determination from scaling
violations is that the gluon only couples strongly to other PDFs
for sufficiently small
$x$: specifically, at large $x$, 
$P_{qg}^S$ in Equation~\ref{dglap} rapidly becomes negligible in comparison to 
$P_{qq}^S$.  Hence, the large $x$ gluon is likely 
affected by large uncertainties, which can only be reduced by looking
at hadronic (jet) final states.

\subsubsection{Global Fits}

In current state of the art ``global'' fits, information on PDFs is
maximized by combining  experimental information on an array of
different physical processes, which provide a handle on different
PDFs or combinations of PDFs, in various kinematic regions. In
Table~\ref{tab:processes}, taken from Reference~\cite{Martin:2009iq},
we list the processes that are included in a typical present-day
global fit (MSTW08), and the PDFs they constrain.
\begin{table}%
  \def~{\hphantom{0}}
  \begin{center}
    \begin{tabular}{llll}
      \hline
      Process & Subprocess & Partons & $x$ range \\ \hline
      $\ell^\pm\,\{p,n\}\to\ell^\pm\,X$ & $\gamma^*q\to q$ & $q,\bar{q},g$ & $x\gtrsim 0.01$ \\
      $\ell^\pm\,n/p\to\ell^\pm\,X$ & $\gamma^*\,d/u\to d/u$ & $d/u$ & $x\gtrsim 0.01$ \\
      $pp\to \mu^+\mu^-\,X$ & $u\bar{u},d\bar{d}\to\gamma^*$ & $\bar{q}$ & $0.015\lesssim x\lesssim 0.35$ \\
      $pn/pp\to \mu^+\mu^-\,X$ & $(u\bar{d})/(u\bar{u})\to \gamma^*$ & $\bar{d}/\bar{u}$ & $0.015\lesssim x\lesssim 0.35$ \\
      $\nu (\bar{\nu})\,N \to \mu^-(\mu^+)\,X$ & $W^*q\to q^\prime$ & $q,\bar{q}$ & $0.01 \lesssim x \lesssim 0.5$ \\
      $\nu\,N \to \mu^-\mu^+\,X$ & $W^*s\to c$ & $s$ & $0.01\lesssim x\lesssim 0.2$ \\
      $\bar{\nu}\,N \to \mu^+\mu^-\,X$ & $W^*\bar{s}\to\bar{c}$ & $\bar{s}$ & $0.01\lesssim x\lesssim 0.2$ \\ \hline
      $e^\pm\,p \to e^\pm\,X$ & $\gamma^*q\to q$ & $g,q,\bar{q}$ & $0.0001\lesssim x\lesssim 0.1$ \\
      $e^+\,p \to \bar{\nu}\,X$ & $W^+\,\{d,s\}\to \{u,c\}$ & $d,s$ & $x\gtrsim 0.01$ \\
      $e^\pm p\to e^\pm\,c\bar{c}\,X$ & $\gamma^*c\to c$, $\gamma^* g\to c\bar{c}$ & $c$, $g$ & $0.0001\lesssim x\lesssim 0.01$ \\
      $e^\pm p\to\text{jet}+X$ & $\gamma^*g\to q\bar{q}$ & $g$ & $0.01\lesssim x\lesssim 0.1$ \\ \hline
      $p\bar{p}\to \text{jet}+X$ & $gg,qg,qq\to 2j$ & $g,q$ & $0.01\lesssim x\lesssim 0.5$ \\
      $p\bar{p}\to (W^\pm\to\ell^{\pm}\nu)\,X$ & $ud\to W,\bar{u}\bar{d}\to W$ & $u,d,\bar{u},\bar{d}$ & $x\gtrsim 0.05$ \\
      $p\bar{p}\to (Z\to\ell^+\ell^-)\,X$ & $uu,dd\to Z$ & $d$ & $x\gtrsim 0.05$
      \\ \hline
    \end{tabular}
  \end{center}
  \caption{The main processes included in the MSTW 2008 global PDF analysis ordered in three groups: fixed-target experiments, HERA and the Tevatron.  For each process we give an indication of their dominant partonic subprocesses, the primary partons which are probed and the approximate range of $x$ constrained by the data.}
  \label{tab:processes}
\end{table}
The CTEQ and NNPDF global fits, to be discussed below
in Section~\ref{sec:status}, have similar features.

Based on this table and the previous discussion we conclude that:
\begin{itemize}
\item information on the overall shape of quarks and gluons at medium
  $x$, as well as on the isosinglet--isotriplet separation, come from
  fixed-target DIS data on proton and deuterium targets 
  (dominated by $\gamma^*$ exchange);
\item an accurate determination of the behavior of the gluon and
  quark at small $x$  (where it is dominated by the singlet in this
  region) and by individual light flavors at medium $x$ (where NC and
  CC data play a r\^ole in separating individual flavors) is found
  from the very precise HERA NC and CC data;
\item information on the flavor separation at small $x$ comes from
  Tevatron Drell--Yan data  (in particular the $W$ asymmetry, as
  discussed above);
\item the flavor separation at  medium $x$ is mostly controlled by
  the Drell--Yan data for fixed proton and nucleus targets;
\item the total valence component is constrained by the neutrino
  inclusive DIS data;
\item strangeness is controlled by neutrino dimuon data, as well as by
  the interplay of the $W$ and $Z$ production data with lower-scale
  DIS and Drell--Yan data;
\item the large $x$ gluon, only weakly determined by DIS scaling
  violations, is further constrained by Tevatron jet data.
\end{itemize}

%%%%%%%%%%

\section{STATUS OF PDF SETS}
\label{sec:status}

Various fitting groups currently produce general-purpose
sets of PDFs of the nucleon, with most of the
groups having a long history which goes back at least a couple of decades,
as summarized in the introduction. Six of these groups have been
providing regular updates of their PDFs, and here we will discuss
their most recent NNLO sets: MSTW08~\cite{Martin:2009iq},
CT10~\cite{Nadolsky:2012ia},  
NNPDF2.3~\cite{Ball:2012cx}, 
HERAPDF1.5~\cite{HERA:2011}, ABM11~\cite{Alekhin:2012ig} and
JR09~\cite{JimenezDelgado:2008hf}. All of these sets are publicly available
though the standard \textsc{lhapdf} interface~\cite{Whalley:2005nh}, though
CT10 NNLO and HERAPDF1.5 have not been presented in a journal
publication. We will not discuss PDFs which are not available from
\textsc{lhapdf}. Also, we will not discuss PDFs 
for different kinds of targets or with more specialized or limited
goals: PDFs for nuclei
or other hadrons  (such as pions); PDFs partly or entirely determined
based on models of hadrons rather than (or in addition to) data; PDFs
for medium-energy physics which incorporate non-perturbative or
higher-twist effects.

\begin{figure}
\centerline{\includegraphics[width=\textwidth]{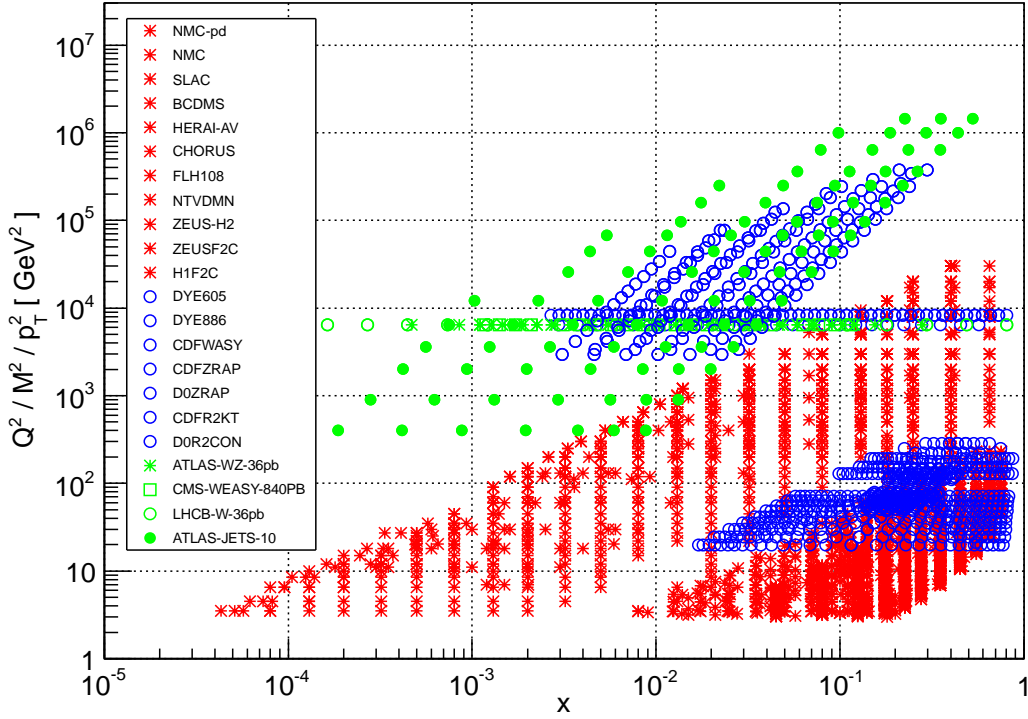}}
\caption{The kinematical coverage of the experimental data used in
  the NNPDF2.3 PDF determination, from Reference~\cite{Ball:2012cx}.}
\label{fig:nnpdf23kin}
\end{figure}

The main feature which distinguishes PDF sets is the data on which
they are based. These are summarized in Table~\ref{tab:compare}.
\begin{table}
  \centering{\scriptsize
  \begin{tabular}{l|c|c|c|c|c|c}
    \hline
    & MSTW08 & CT10 & NNPDF2.3 & HERAPDF1.5 & ABM11 & JR09 \\
    \hline
    { HERA DIS} & \tick & \tick & \tick & \tick & \tick & \tick \\
    { Fixed-target DIS} & \tick & \tick & \tick & \cross & \tick & \tick \\
    { Fixed-target DY} & \tick & \tick & \tick & \cross & \tick & \tick \\
    { Tevatron $W$+$Z$+jets} & \tick & \tick & \tick & \cross & \cross & \cross \\
    { LHC $W$+$Z$+jets} & \cross & \cross & \tick & \cross & \cross & \cross \\
    \hline
  \end{tabular}}
  \caption{Data included in various NNLO PDF sets.}
  \label{tab:compare}
\end{table}
Only three
groups (MSTW08~\cite{Martin:2009iq}, CT10~\cite{Nadolsky:2012ia}, and
NNPDF2.3~\cite{Ball:2012cx}) make a fully global fit, defined
here to be a fit including HERA and fixed-target DIS data,
fixed-target Drell-Yan production, and Tevatron data on $W$, $Z$ and
jet production. The NLO version of the JR09 fit,
GJR08~\cite{Gluck:2007ck}, does include some Tevatron jet data.
The NNPDF2.3 set is the
only one to include LHC data; in order to assess
 the impact of the additional LHC data,
we will also compare to a variant of the NNPDF2.3 analysis
without LHC data.  Concerning HERA data, note that CT10 and NNPDF2.3 
include the combined HERA I inclusive data~\cite{Aaron:2009aa},
 MSTW08 and JR09 instead include the older separate data from H1 and
ZEUS, ABM11 includes combined HERA I data but only with the cut
$Q^2<1000$~GeV$^2$,
and HERAPDF1.5 additionally includes the preliminary
combined HERA II inclusive data~\cite{HERA:2010}. The kinematical
coverage of the
NNPDF2.3 data set is shown in Figure~\ref{fig:nnpdf23kin}, with the $x$
and $Q^2$ values shown determined using leading-order parton kinematics.

As discussed in Section~\ref{sec:det}, various alternative choices are
possible in PDF determination both in terms of theory and
methodology. The main choices which underlie the PDF sets we consider
here are summarized in Table~\ref{tab:comparth}.
\begin{table}
  \centering{\scriptsize
  \begin{tabular}{l|c|c|c|c|c|c}
    \hline
    &  MSTW08 &  CT10 & NNPDF2.3 &  HERAPDF1.5 & ABM11 & JR09 \\
    \hline
    {No.~of PDFs} & 7 & 6 & 7 & 5 & 6 & 5 \\
    {Statistics} & Hess.+DT & Hess.+DT & MC & Hess.+Model+Parm. & Hess. & Hess.+T \\
    {PDF parms.} & 20+8 & 25 & 259 & 14 & 24 & 12 \\
    {Heavy quarks} & VFN TR & VFN ACOT  & VFN FONLL & VFN TR & FFN & FFN \\
    \hline
  \end{tabular}}
  \caption{Main features of various NNLO PDF sets (see text for details).}
  \label{tab:comparth}
\end{table}
All sets are now
available at NLO and NNLO, and all but HERAPDF also have a LO version,
though, as mentioned in Section~\ref{sec:pert}, LO PDFs are often
optimized for use with Monte Carlo event generators, and thus we will
not discuss them further. 
 The number of independently parametrized
PDFs varies between seven (the three lighter quarks and antiquarks and
the gluon), six (the total strangeness is independently parametrized,
but not the strange and antistrange separately) or five (strangeness
not fitted, and
assumed to be a fixed fraction of the sea). The methodology for
uncertainty representation and determination is Hessian, based on a
parametrization of the form of Equation~\ref{pdfparmgen} for all groups
except NNPDF, which uses a Monte Carlo representation based on a
neural network parametrization (see Section~\ref{sec:fit}). In order to
determine confidence levels, CT and
MSTW use dynamical tolerance, denoted in the table as ``DT''
(recall Section~\ref{sec:fit}), JR use simple
tolerance, denoted by ``T'', HERAPDF uses $\Delta
\chi^2=1$ but supplemented by an estimate of model
and parametrization uncertainties, and
ABM just use $\Delta\chi^2=1$. In each case
the total number of free parameters at NNLO is as given in the table;
the number of parameters at NLO is the same for all groups but CT10,
which at NLO has 26 parameters, and HERAPDF1.5, which at NLO has only 10
parameters.  MSTW08 uses 28 free parameters for the determination
of the best fit, 8 of which are fixed when determining uncertainties.
JR09 introduces the further ``dynamical'' assumption that PDFs
are valence-like at a low scale  $Q_0^2<1$~GeV$^2$.
All groups but ABM and JR use
variable-flavor number schemes, with heavy-quark masses included using
one of the matching methods discussed in Section~\ref{sec:th}. The
treatment of $\alpha_S$ will be discussed in more detail below.

\subsection{Values and Uncertainties of Strong Coupling}
\label{sec:aval}

An important issue which distinguishes PDF sets is the treatment
of the strong coupling $\alpha_S(M_Z^2)$.  Because the value of
$\alpha_S$ is strongly correlated with PDFs, one should always use in
cross-section calculations PDFs which have been determined with the
same value of $\alpha_S$ that is adopted for the calculation itself.  

The value of $\alpha_S(M_Z^2)$, and its uncertainty, can either be determined
simultaneously with the PDFs or imposed as an external constraint.
Furthermore, if the value of $\alpha_S$ is determined simultaneously
with the PDFs, the quoted value of the PDF uncertainties may refer strictly to
the PDF-only uncertainty as $\alpha_S$ is kept fixed at its best-fit value,
or it may also include the uncertainty due to the variation of $\alpha_S$
itself.

\begin{figure}
  \centerline{\includegraphics[width=\textwidth]{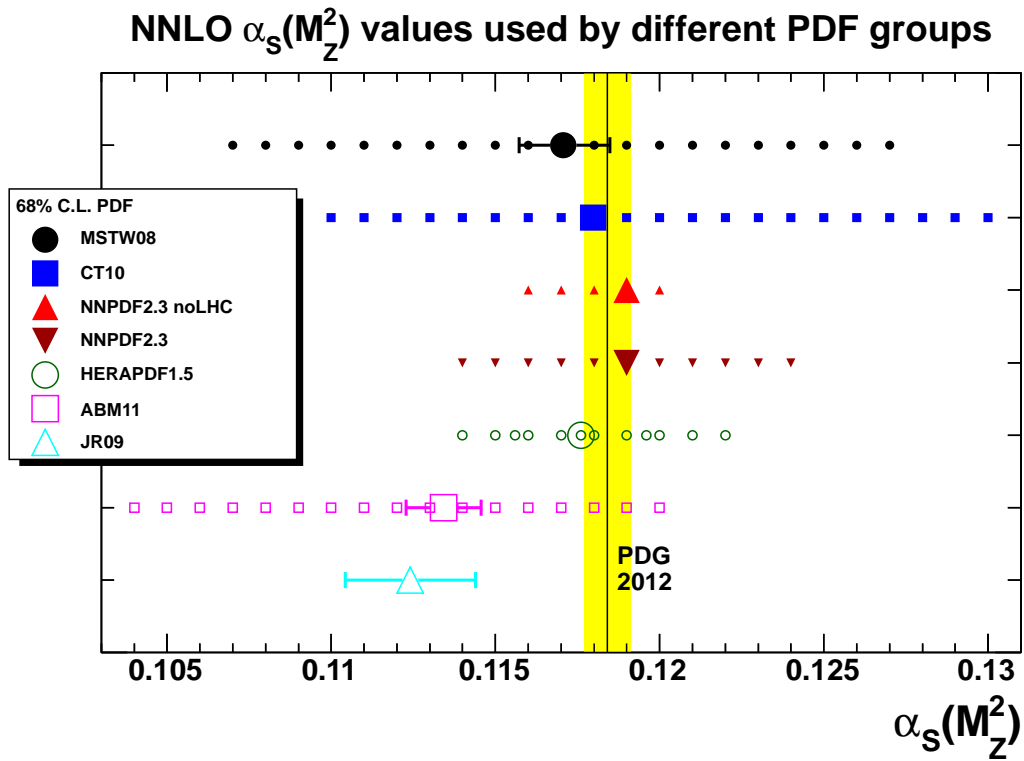}}
  \caption{$\alpha_S(M_Z^2)$ values for which NNLO PDFs are provided
    by various groups. The larger symbols denote the values used in
    subsequent plots.}
  \label{fig:alphaSMZ}
\end{figure}

The values of $\alpha_S(M_Z^2)$ used by different NNLO PDF fitting
groups are shown in Figure~\ref{fig:alphaSMZ}, where the larger
symbols represent the default value used by each group, that is used
for the determination of PDF uncertainties.
For MSTW08, ABM11 and JR09, this value is determined from the fit with
uncertainties shown by the horizontal error bars, while for CT it is
chosen as a fixed value close to the PDG world
average~\cite{Beringer:1900zz}, also shown in the plot. NNPDF do not
have a default value and provide a full Monte Carlo replica set for
each of the $\alpha_S$ values shown, though they have also presented an
$\alpha_S$ determination~\cite{Ball:2011us} 
based on their previous NNPDF2.1 set, with
results consistent with the PDG average.
For NNPDF, which does not have a
default value, we arbitrarily choose
$\alpha_S(M_Z^2)=0.119$ as default to be used in all plots, so that the MSTW08
and NNPDF $\alpha_S$ values bracket the CT10 value by providing a
variation of $\Delta\alpha_S=0.001$ about it, for reasons to be discussed in
Section~\ref{sec:combination} below.

The smaller symbols in Figure~\ref{fig:alphaSMZ} indicate
the PDF sets with alternative values of $\alpha_S(M_Z^2)$ provided by
each group.  All groups provide only the  best-fit PDF set
for each of these values, except NNPDF which instead provide
a full set for each value. 
The PDF uncertainties provided by MSTW and CT
at the reference value of $\alpha_S$, and by NNPDF for all values of
$\alpha_S$, do not include the $\alpha_S$ uncertainty, though MSTW
also provide additional sets allowing combined PDF+$\alpha_S$
uncertainties~\cite{Martin:2009bu}.  JR and ABM only
provide combined PDF+$\alpha_S$ uncertainties.

\subsection{Comparison of PDFs}
\label{sec:comp}

\begin{figure}
  \centerline{\includegraphics[width=\textwidth]{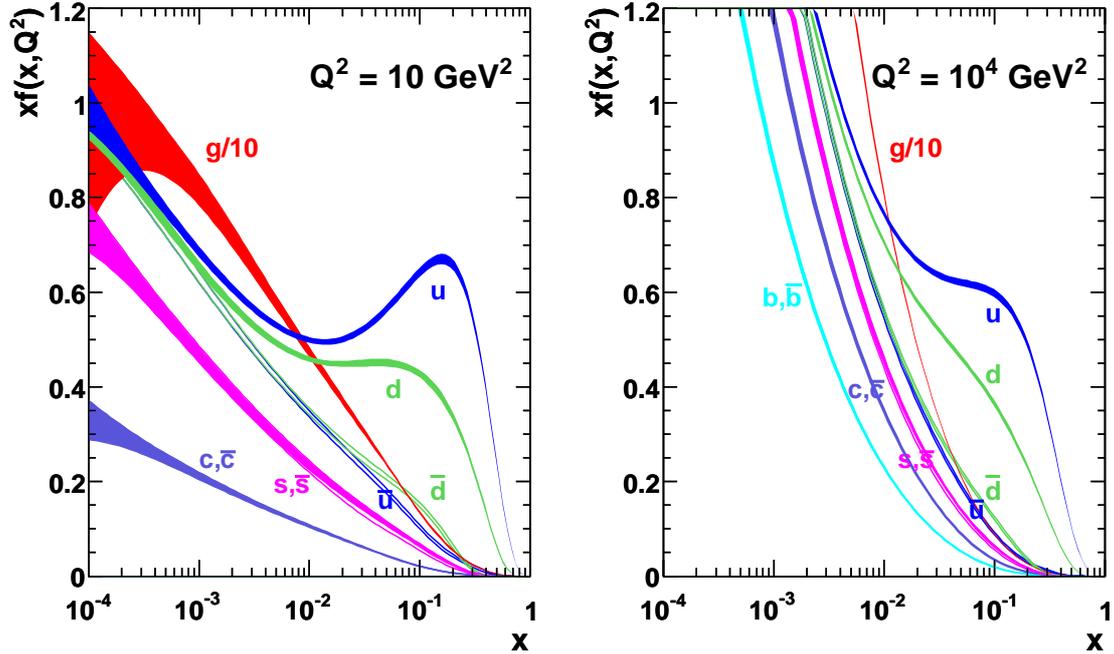}}
  \caption{MSTW 2008 NNLO PDFs at $Q^2=10$~GeV$^2$ and $Q^2=10^4$~GeV$^2$,
    from Reference~\cite{Martin:2009iq}.}
  \label{fig:allpdfs}
\end{figure}

A typical set of PDFs (MSTW08) at two different scales is
shown in Figure~\ref{fig:allpdfs}.
It is clear from Equation~\ref{hadrfact}, however, that, in order
to understand properties of hadronic cross sections and the impact of
PDFs on them, it is more useful to consider the relevant parton--parton
luminosities, defined as in Equation~\ref{lumdef}, rather than the
PDFs themselves.

A detailed study would involve comparison of the luminosity for all
parton combinations: for example, $\mathcal{L}_{u\bar{d}}$, relevant
for  $W^+$ production, and so forth. Here we only compare the $q\bar q$
luminosity summed over quark flavors, i.e.,
\begin{equation}\label{lumqqbdef}
\sum_{q=u,d,s,c,b} \left(\mathcal{L}_{q\bar{q}}+\mathcal{L}_{\bar{q}q}\right),
\end{equation}
where $\mathcal{L}_{ab}$ is defined in Equation~\ref{lumdef}.
\begin{figure}
  (a)\\
  \centerline{\includegraphics[width=0.94\textwidth]{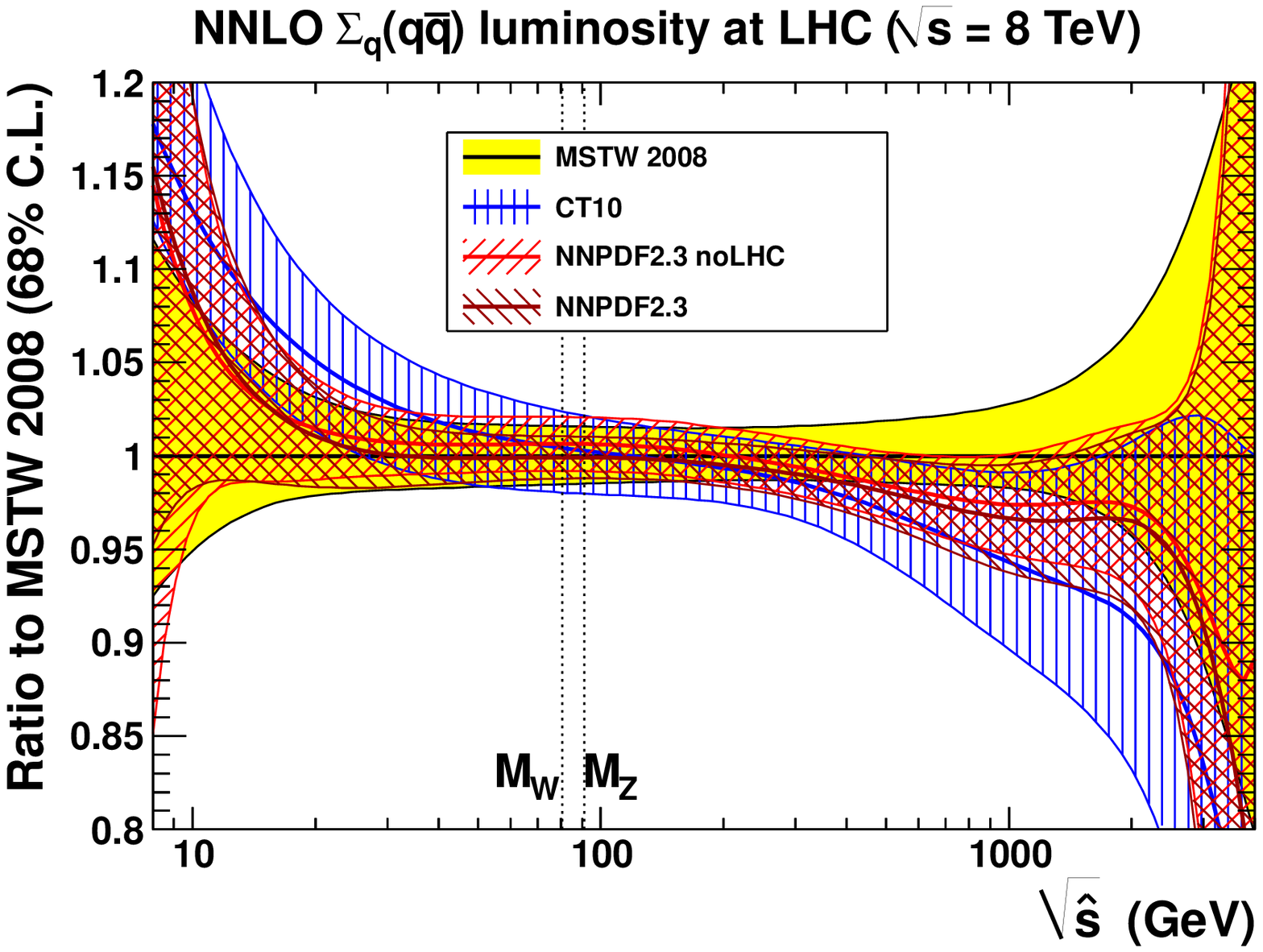}}
  (b)\\
  \centerline{\includegraphics[width=0.94\textwidth]{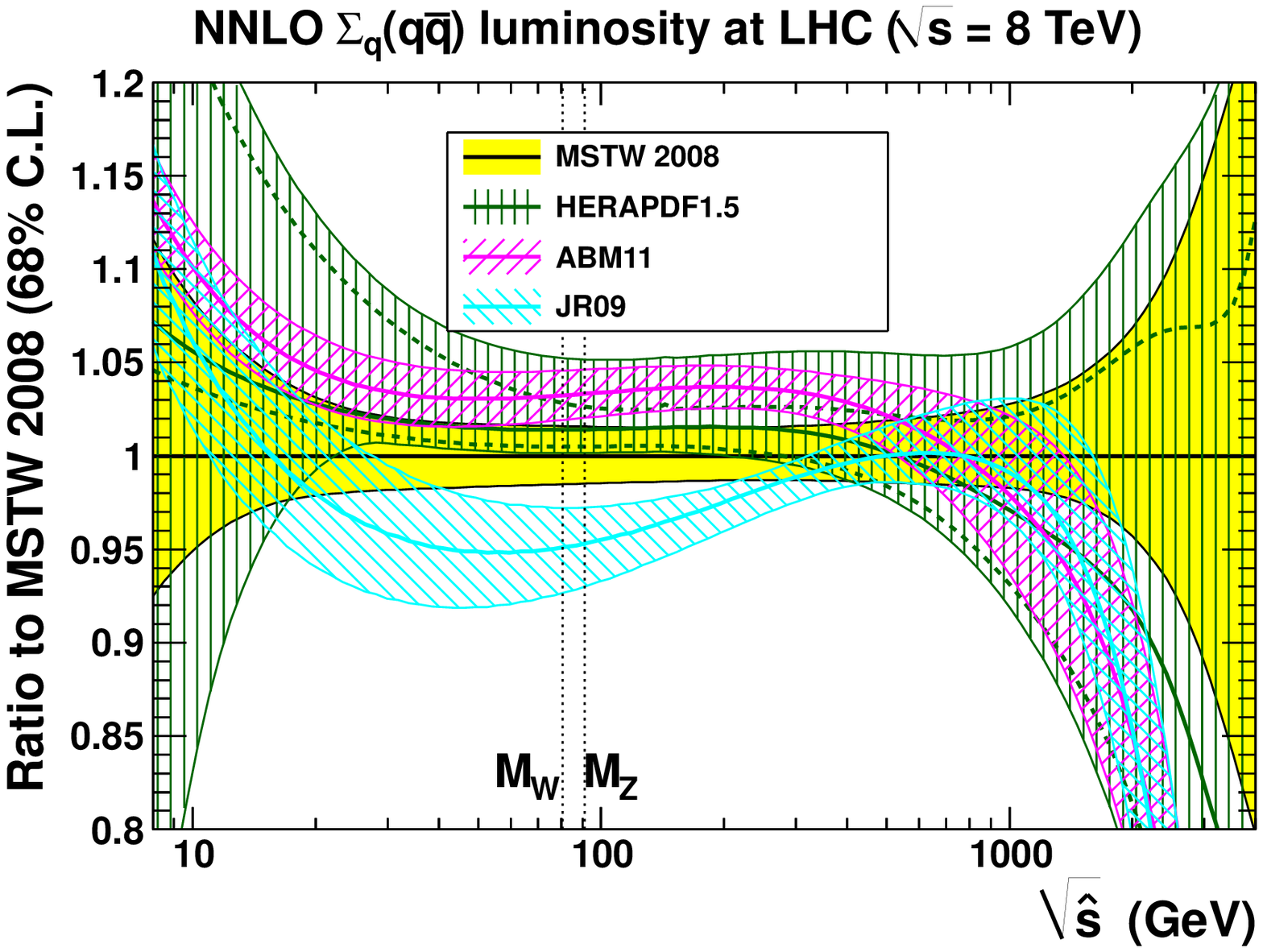}}
  \caption{NNLO $q\bar{q}$ luminosity functions taken as the ratio to MSTW08.  (a)~MSTW08 vs.~CT10 vs.~NNPDF2.3noLHC vs.~NNPDF2.3, then (b)~MSTW08 vs.~ABM11 vs.~HERAPDF1.5 vs.~JR09.}
\label{fig:qqbarlumi}
\end{figure}
\begin{figure}
  (a)\\
  \centerline{\includegraphics[width=0.94\textwidth]{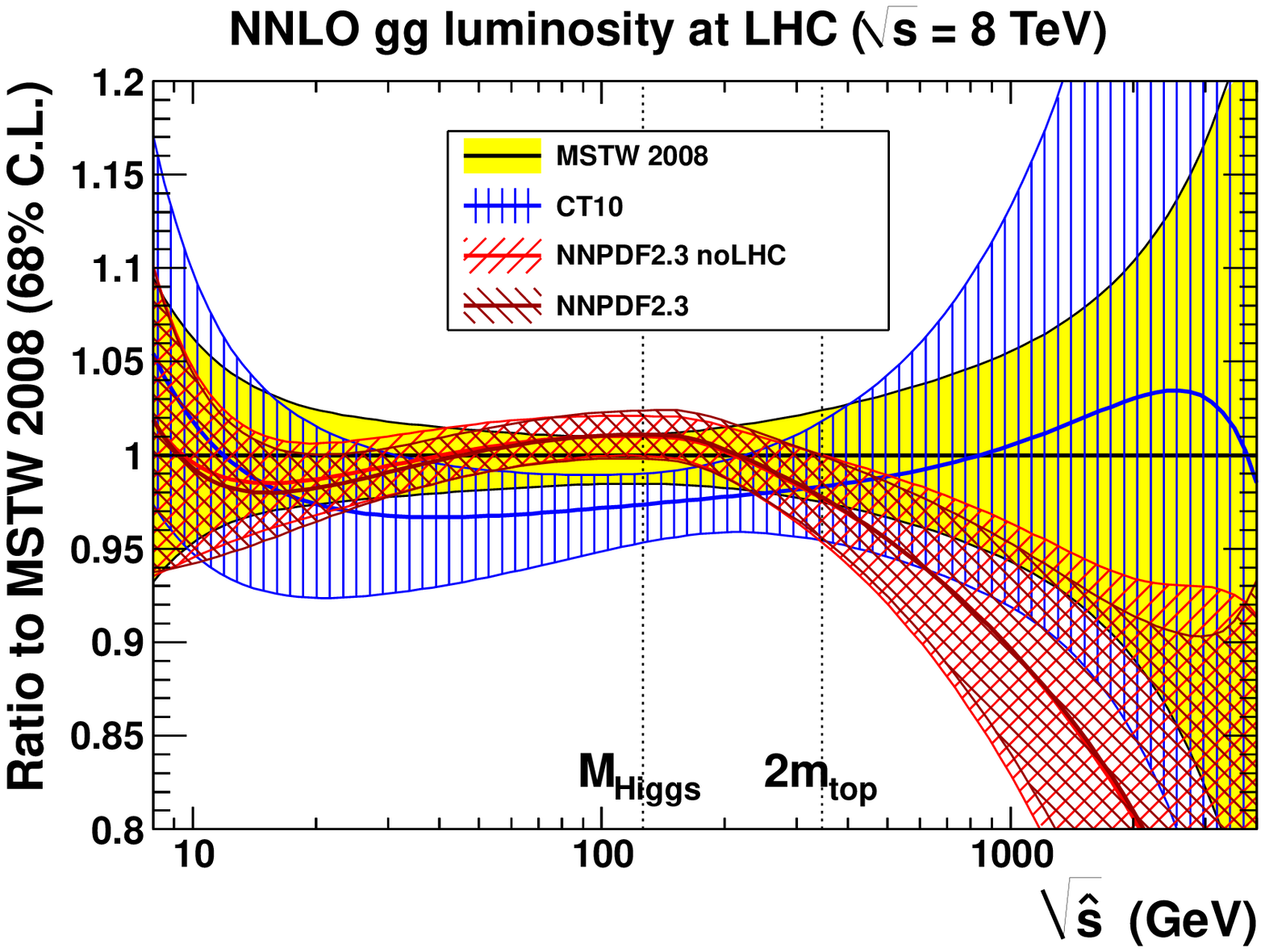}}
  (b)\\
  \centerline{\includegraphics[width=0.94\textwidth]{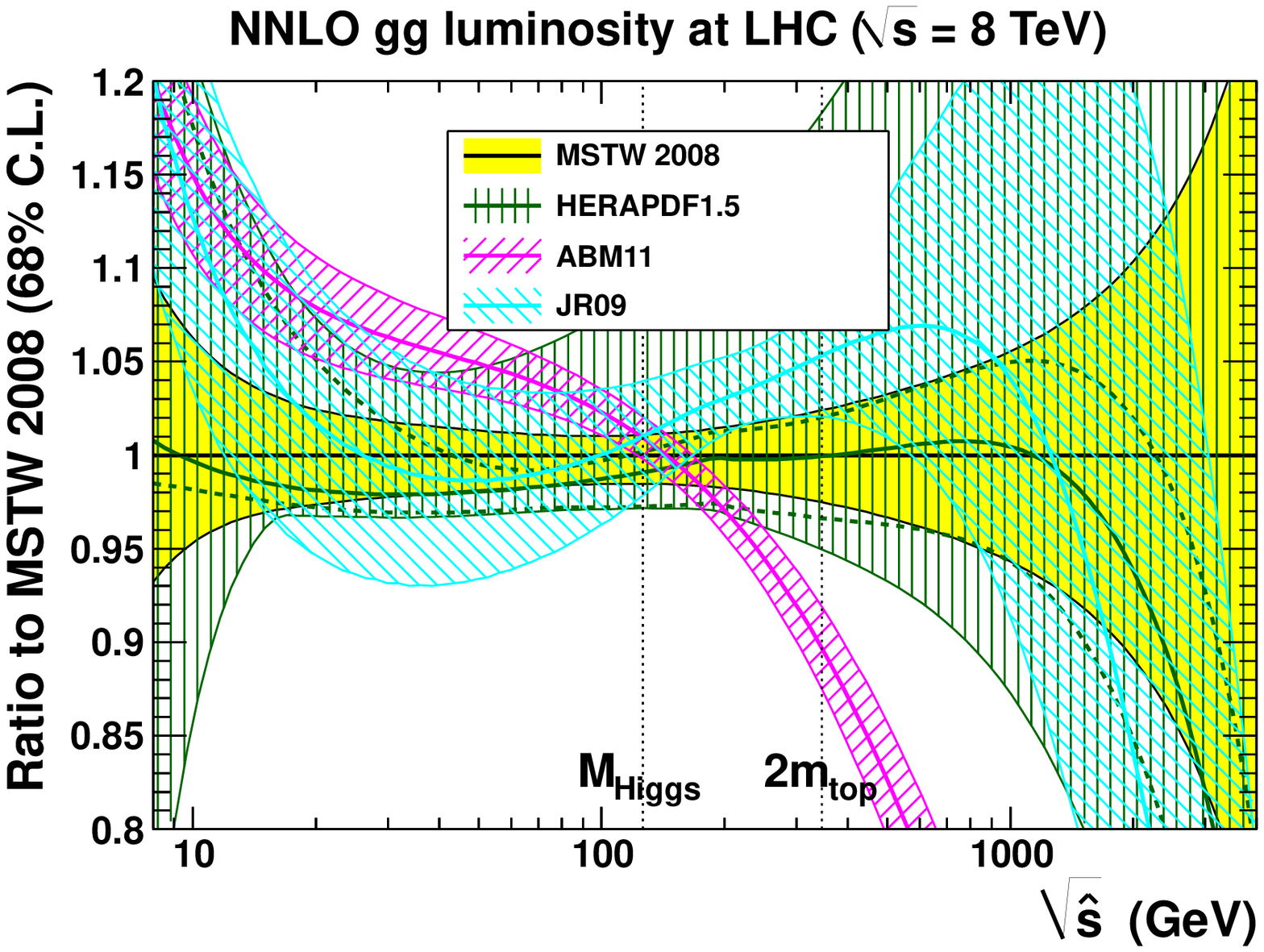}}
  \caption{NNLO $gg$ luminosity functions taken as the ratio to MSTW08.  (a)~MSTW08 vs.~CT10 vs.~NNPDF2.3noLHC vs.~NNPDF2.3, then (b)~MSTW08 vs.~ABM11 vs.~HERAPDF1.5 vs.~JR09.}
  \label{fig:gglumi}
\end{figure}
In Figure~\ref{fig:qqbarlumi} and Figure~\ref{fig:gglumi} 
we show, respectively, the NNLO $q\bar{q}$  and $gg$ luminosities,
displayed as a ratio to the MSTW 2008 NNLO
luminosities, for the LHC at $\sqrt{s} = 8$~TeV, and in
Figure~\ref{fig:gglumi} we show the corresponding NNLO $gg$
luminosities.  We use the  $\alpha_S$ values for each set
shown in Figure~\ref{fig:alphaSMZ}.
Note that all uncertainty bands
are shown at 68\% C.L., requiring the CT10 uncertainties
(corresponding to a nominal 90\% C.L.) to be divided by a factor of
1.64485. Similar plots, using a common value of $\alpha_S(M_Z^2)=0.118$,
can be found in Reference~\cite{Ball:2012wy}.

The relevant values of $\sqrt{\hat{s}} = M_{W,Z}$ are indicated for
the $q\bar{q}$ luminosities, and the relevant values of
$\sqrt{\hat{s}} = M_H,2m_t$ (for $M_H = 126$~GeV and $m_t=173.18$~GeV)
are indicated for the $gg$ luminosities.  There is fairly good
agreement for the three global fits (MSTW08, CT10 and NNPDF2.3), but
more variation for the other sets, which confirms that the dominant
factor in determining the features of the PDFs is the choice of data set.
There is little difference between the luminosities computed using
NNPDF2.3 and NNPDF2.3noLHC, which shows that the impact of the LHC
data is moderate. We will see in Section~\ref{sec:pheno} that this is
often but not always the case.
 The NLO trend between groups is
similar  to the NNLO trend, with the exception of HERAPDF at large $\hat{s}$
values, where the HERAPDF1.5 NLO set~\cite{HERA:2010} has a much
larger $q\bar{q}$
luminosity, and a much softer $gg$ luminosity, than other NLO PDF
groups. 

These luminosities are the basic input to LHC phenomenology, as we will
discuss shortly. Current recommendations~\cite{Botje:2011sn} to use
global fits for LHC searches and calibration,  already
mentioned in Section~\ref{sec:intro} and to be discussed in
Section~\ref{sec:pheno} below, were based on similar, more detailed
comparisons of luminosities and PDFs made in 2010~\cite{Alekhin:2011sk}.
However, the situation is much improved now than in 2010, when
only MSTW08 had a NNLO PDF set from a global fit, and 
differences at NLO between MSTW08, CTEQ6.6 and NNPDF2.0 were larger,
due to, for example, the use of a less flexible
gluon parametrization in CTEQ6.6
and the lack of inclusion of terms suppressed by powers of the
charm-quark mass in NNPDF2.0. 

A recent thorough analysis of PDFs and
luminosities~\cite{Ball:2012wy} 
shows that the general features of NNLO global PDF sets, at a scale of
order of $Q^2\sim M_W^2$,  are the following,
bearing in mind that experimental information is not available outside
the region $10^{-4}\lesssim x\lesssim 0.4$ (see
Figure~\ref{fig:nnpdf23kin}). 
Up and down quark and antiquark 
distributions are known
to an accuracy better than about $5\%$ in a wide range of $x$,
roughly $10^{-4}\lesssim
x\lesssim 0.3$ for the up distribution, $10^{-4}\lesssim
x\lesssim 0.1$ for the down and the antiup distribution, $10^{-4}\lesssim
x\lesssim 0.01$ for the antidown distribution,
and there is good agreement between the three global
sets.  For smaller values of $x$, uncertainties gradually blow up,
but there remains good
agreement between sets as the behavior in this region is mostly
driven by perturbative evolution, while, for larger values of $x$,
uncertainties blow up and widely different behaviors are observed
between sets: already for $x\sim0.5$ uncertainties are likely to be
bigger than $10\%$, and perhaps underestimated  especially as $x$ grows.
Strangeness is nominally known to about 10--15\% accuracy in the region
$0.003\lesssim x \lesssim 0.1$. However, it should be kept in mind
that strangeness is largely determined by neutrino dimuon data (see
Section~\ref{strangesec}), which are subject to various poorly
controlled systematics, and also, one of the three global sets does not
parametrize independently the $s$ and $\bar s$ distribution, while
another only has a small number of parameters. Indeed, disagreement
between different sets are up to the 30\% level. The gluon
distribution is known with an accuracy which is comparable or
marginally worse than  that of
light quarks, i.e.,~$\sim5\%$ at small $10^{-4}\lesssim x\lesssim 0.1$, but
rapidly deteriorates at larger $x$, where it is only constrained by
jet data. As already mentioned, here the agreement between global sets
is not as good as one might hope, and discrepancies up to the level of
1.5--2 sigma between global fits are observed in the region $x\sim
0.02$, which is relevant for Higgs production.

Comparison of NLO and NNLO PDFs suggests that uncertainties related to
higher-order corrections are smaller than $5\%$ in the region where
PDFs are currently determined, meaning that the neglected
theory uncertainties are likely to be smaller than the
experimental PDF uncertainties at NNLO, whereas at NLO they might be comparable.

%%%%%%%%%%

\section{LHC PHENOMENOLOGY}
\label{sec:pheno}

The first LHC proton run was completed in December 2012, after a
remarkable three years, with a center-of-mass energy of 7~TeV in
2010/2011 and 8~TeV in 2012.  Data have been collected for a vast
array of Standard Model processes. Many of these are already
leading to new significant constraints on PDFs, with others holding the
promise to do so in the very near future, and the knowledge of PDFs
has played a significant r\^ole in the discovery of a Higgs-like
particle~\cite{Aad:2012gk,Chatrchyan:2012gu}.  In this section, we confront LHC
data with the predictions of various PDF sets for some key Standard
Model total cross sections, specifically $W$, $Z$, Higgs boson and
top-pair production, then we discuss methods for combining the
predictions made using the PDF sets from different groups.  We finally
assess the current constraints on PDFs provided by LHC data and we
examine the prospects for future improvements.

\subsection{Predictions for LHC Cross Sections}

A comprehensive study of the PDF dependence of key LHC cross sections
has recently been made in Reference~\cite{Ball:2012wy}, following
earlier work in
References~\cite{Alekhin:2011sk,Watt:2011kp,Watt:2012np}.  Here we
will present some selected results from a continuation of the earlier
study~\cite{Alekhin:2011sk,Watt:2011kp,Watt:2012np}, but now updated
to account for the latest PDF sets and LHC data.  
The impact of LHC data will be discussed in more detail in
Section~\ref{sec:constraints}, but we will already show here results
with the two variants of NNPDF2.3, with and without LHC data, which
will give us an indication of their current impact.

\subsubsection{$W$ and $Z$ Production}

The $W^+$, $W^-$ and $Z$ cross sections at the LHC differ from those at
the Tevatron discussed in Section~\ref{sec:data} because the LHC is a
$pp$ rather than a $p\bar p$ collider. Hence, somewhat different
combinations of the light quark and antiquark distributions are
measured. To understand this, it is useful to consider the 
$W^\pm\equiv W^++W^-$ to $Z^0$  cross-section ratio, and the $W^+$
to $W^-$ cross-section ratio. Assuming that very roughly
$\bar{u}(x_2)\approx\bar{d}(x_2)$, and neglecting heavier quarks, one gets
\begin{align}
  \label{eq:WoverZ}
  & \frac{\sigma_{W^+}+\sigma_{W^-}}{\sigma_{Z^0}}\sim
  \frac{u(x_1)+d(x_1)}{0.29\,u({x}_1)+0.37\,d({x}_1)},\\
  \label{WpWm}
  & \frac{\sigma_{W^+}}{\sigma_{W^-}}\sim\frac{u(x_1)\,\bar{d}(x_2)}
  {d(x_1)\,\bar{u}(x_2)}\sim\frac{u(x_1)}{d(x_1)},
\end{align}
where $x_1$ and $x_2$ are fixed when measuring a
rapidity distribution, and
we have assumed that $x_{1,2}$ are in a region where
$q(x_1)\bar{q}(x_2)$
dominates over $\bar{q}(x_1)q(x_2)$. Equation~\ref{eq:WoverZ}
shows that the $W^\pm$
and $Z$ cross sections are very highly correlated (so that their ratio
depends very little on the PDFs), while Equation~\ref{eq:WoverZ} (to
be compared to its counterpart at a $p\bar p$ collider,
Equation~\ref{dywasym}) shows that the ratio of $W^+$/$W^-$ cross sections is a
sensitive probe of the $u/d$ ratio.

We now consider explicitly predictions for $W$ and $Z$ production,
where, for definiteness, we show results only for total
cross sections. However, as mentioned, only rapidity distributions
probe fixed leading-order parton kinematics.  Furthermore,
only around half of the total $W^\pm$ and $Z^0$ cross sections lie
inside the acceptance of the ATLAS and CMS detectors, so that a theory
calculation is needed to extrapolate the measurement over the whole phase space,
introducing an additional uncertainty on the total cross sections.  
Therefore, data-to-theory
comparisons for precision physics are best made at the level of
the fiducial cross
section (i.e., within the acceptance), which is possible at NNLO using
the public \textsc{fewz}~\cite{Gavin:2010az,Gavin:2012sy}
and \textsc{dynnlo}~\cite{Catani:2009sm} codes, and indeed was done in
the ATLAS publication~\cite{Aad:2011dm}.

\begin{figure}
  (a)\\
  \centerline{\includegraphics[width=0.66\textwidth]{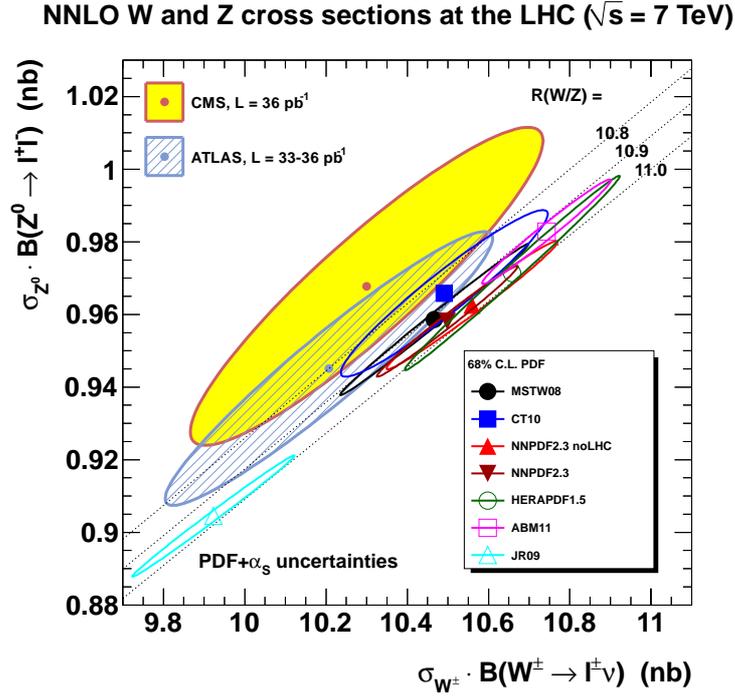}}
  (b)\\
  \centerline{\includegraphics[width=0.66\textwidth]{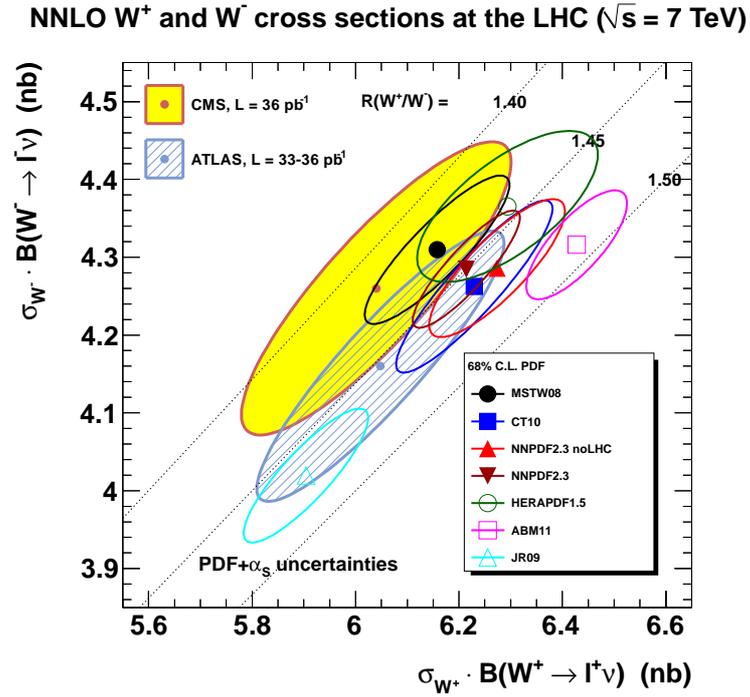}}
  \caption{(a)~$W^\pm$ vs.~$Z^0$ and (b)~$W^+$ vs.~$W^-$ total cross sections
    at NNLO, compared to data from CMS~\cite{CMS:2011aa} and
    ATLAS~\cite{Aad:2011dm}. Lines (dotted) of constant
    ratio are also drawn for reference.}
  \label{fig:wz}
\end{figure}
In Figure~\ref{fig:wz} we show $W^\pm\equiv W^++W^-$ versus $Z^0$ and
$W^+$ versus $W^-$ total cross sections. 
 We also compare to the experimental measurements using the
2010 LHC data from ATLAS~\cite{Aad:2011dm} and CMS~\cite{CMS:2011aa}.
The measured $Z^0$ cross sections have been
corrected~\cite{Watt:2011kp} for the small $\gamma^*$ contribution and
the finite invariant-mass range of the lepton pair (different for
ATLAS and CMS) using a theory calculation at
NNLO~\cite{Anastasiou:2003ds}.  The ellipses are drawn to account for
the correlations between the two cross sections, both for the
experimental measurements and for the theoretical predictions, in such
a way that the projection onto either axis
gives the one-sigma uncertainty for the individual cross sections, so
that the area of the two-dimensional ellipse corresponds to a
confidence-level somewhat smaller than the conventional
68\%~\cite{cowan}.  

The spread in predictions using the different PDF sets is comparable to
the (dominant) luminosity uncertainty of 4\% (CMS) or 3.4\% (ATLAS),
with the JR09 prediction being a clear outlier. The correlation of $W^\pm$ and
$Z$ cross sections is clearly visible
from the plot. The impact of LHC data can be gauged by comparing the
NNPDF2.3noLHC and NNPDF2.3 predictions: while for the $W^\pm$ and $Z$
cross sections there is essentially no difference, consistent with
the stability of the $q\bar q$ luminosity of
Figure~\ref{fig:qqbarlumi}, there is a clear reduction of uncertainty
in the $W^+/W^-$ cross-section ratio, seen as a shrinking of the
corresponding ellipse, which comes from an improved knowledge of the
light flavor separation.

\subsubsection{Higgs and Top-pair Production}
\label{sec:ttbar&ggH}

Whereas the cross sections for production of $W$ and $Z$ bosons are
sensitive to the quark  distributions, we now turn to processes that
are sensitive to the gluon distribution.  The dominant
production mechanism for both Standard Model Higgs bosons, or top-pairs,
at the LHC is through gluon--gluon fusion.  The $gg\to H$ process proceeds
mainly through a top-quark
loop, so both processes start at $\mathcal{O}(\alpha_S^2)$ at LO and 
 are directly sensitive to the value of
$\alpha_S(M_Z^2)$.  Indeed, the CMS measurement of the $t\bar{t}$
cross section has even been used to extract
$\alpha_S$~\cite{CMS:2012sxa}.
Moreover, the gluon PDF itself, being necessarily determined
through strong-interaction processes, is the most sensitive to the
value of $\alpha_S$. Therefore, for these observables we will present
predictions for cross sections as a function of $\alpha_S(M_Z^2)$.

\begin{figure}
  (a)\\
  \centerline{\includegraphics[width=0.94\textwidth]{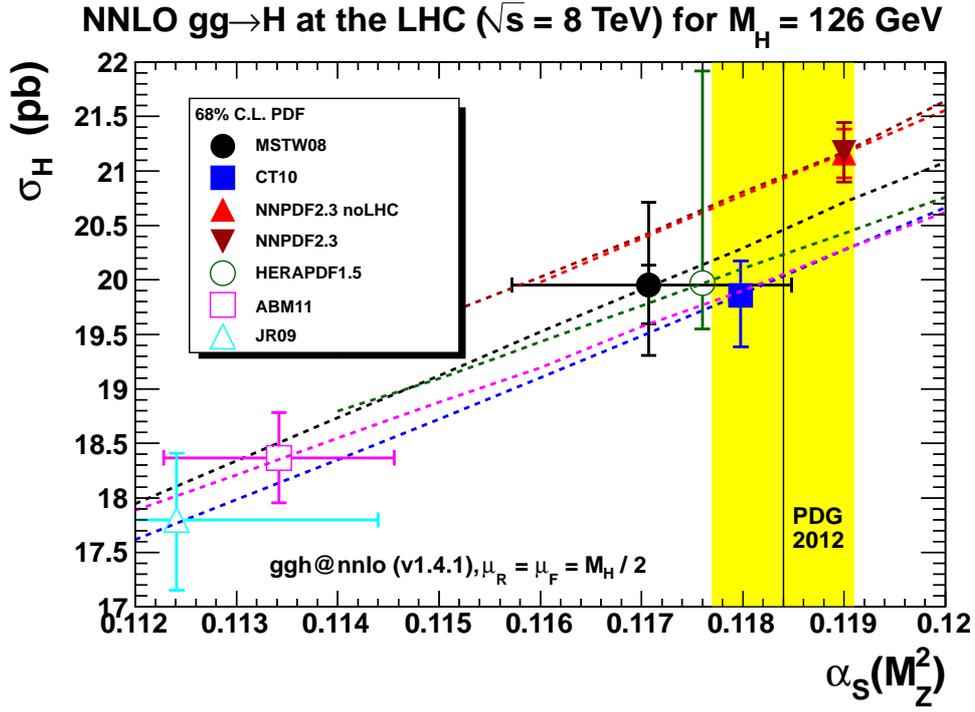}}
  (b)\\
  \centerline{\includegraphics[width=0.94\textwidth]{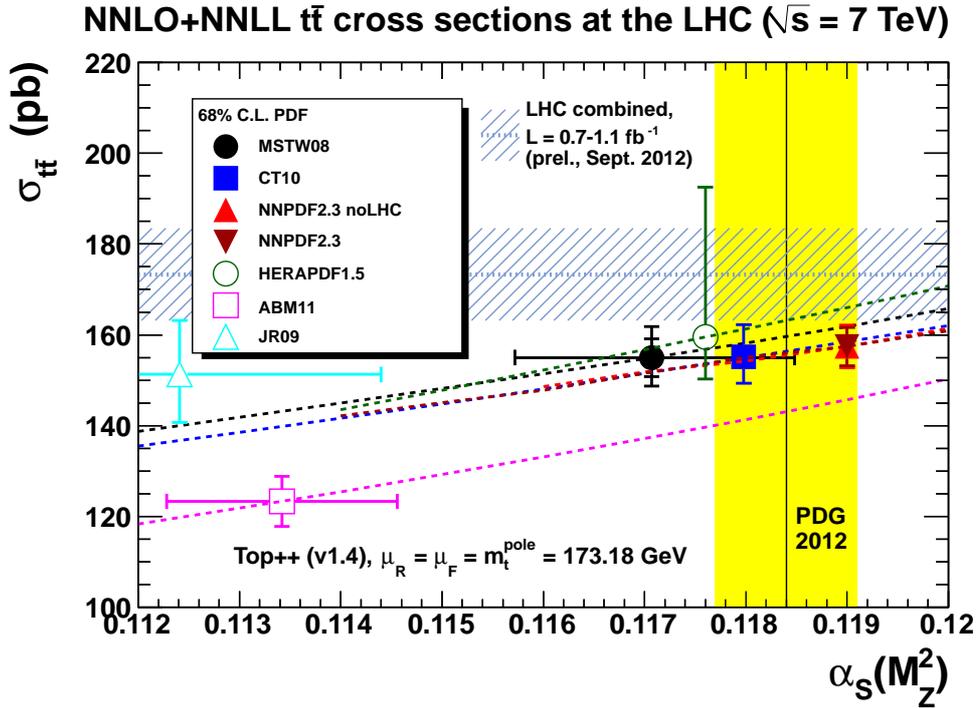}}
  \caption{(a)~NNLO $gg\to H$ total cross sections for $M_H=126$~GeV,
    and (b)~NNLO$_{\rm approx.}$+NNLL $t\bar{t}$ total cross sections
    for $m_t=173.18$~GeV, both plotted as a function of $\alpha_S(M_Z^2)$.}
  \label{fig:ttbar&ggH}
\end{figure}
The $gg\to H$ and $t\bar{t}$ cross sections are shown in
Figure~\ref{fig:ttbar&ggH}
for a Higgs mass $M_H=126$~GeV and a top-quark pole mass of
$m_t=173.18$~GeV~\cite{Aaltonen:2012ra}, for the LHC at $8$~TeV and
$7$~TeV, respectively, probing the gluon distribution at different
$x\sim M_H/\sqrt{s}=0.02$ and $x\gtrsim 2m_t/\sqrt{s}=0.05$.
For Higgs production we use
the \textsc{ggh@nnlo} (version 1.4.1) code~\cite{Harlander:2002wh}
with a scale choice of $\mu_R=\mu_F=M_H/2$, while for top-pair
production we use the \textsc{top++} (version 1.4)
code~\cite{Czakon:2012pz} with a scale choice of $\mu_R=\mu_F=m_t$.
Higgs production at NNLO, in the limit of a heavy top-quark mass, was
originally calculated in
References~\cite{Harlander:2002wh,Anastasiou:2002yz,Ravindran:2003um}.
The top-pair calculations~\cite{Czakon:2012pz} include exact NNLO
corrections for all quark-initiated processes (and $qg\to t\bar{t}$),
with approximate NNLO for $gg\to t\bar{t}$, together with soft-gluon
resummation to next-to-next-to-leading logarithmic accuracy.  The
markers in Figure~\ref{fig:ttbar&ggH} are centered on the
$\alpha_S(M_Z^2)$ values of Figure~\ref{fig:alphaSMZ} and the
corresponding predicted cross section
of each PDF fitting group.  The horizontal error bars span the
$\alpha_S(M_Z^2)$ uncertainty, while the vertical error bars span the
PDF uncertainty, which (recall Section~\ref{sec:aval}) for ABM11 and
JR09 necessarily correspond to a combined PDF+$\alpha_S$ uncertainty.  
For MSTW08, the inner vertical error bars span the PDF only
uncertainty and the outer vertical error bars span the
PDF+$\alpha_S$ uncertainty.  

The $\alpha_S$ dependence of results is shown from the dashed lines,
which interpolate the cross-section predictions calculated using the
sets with different $\alpha_S(M_Z^2)$ values provided by each group.  The
vertical shaded band indicates the PDG world average
value~\cite{Beringer:1900zz} of $\alpha_S(M_Z^2)$, while the
horizontal shaded band in Figure~\ref{fig:ttbar&ggH}(b) indicates the
preliminary combination of ATLAS and CMS $t\bar{t}$ cross-section
measurements~\cite{ATLAS:2012dpa,CMS:2012dya}.  The scale dependence,
obtained by varying $\mu_R$ and $\mu_F$ by factors of two, subject to
the constraint $1/2\le \mu_R/\mu_F\le 2$, leads to a theoretical
uncertainty of $^{+5.6}_{-4.9}\%$, while the top-mass dependence
obtained by varying $m_t=173.18\pm0.94$~GeV~\cite{Aaltonen:2012ra}
leads to an uncertainty on the predicted cross section of $\pm 2.8\%$,
where these percentage uncertainties were obtained using the central
MSTW08 PDF set.  

The strong dependence on the value of $\alpha_S$ is clearly seen. For
the Higgs cross section it is interesting to observe (comparing also
the $gg$ luminosity plot in Figure~\ref{fig:gglumi}) that the value of
$M_H\sim126$~GeV with $\sqrt{s}=8$~TeV is especially unlucky, in that
predictions obtained using
global fits maximally disagree. Note that the NNPDF2.3 and
NNPDF2.3noLHC in these plots are essentially identical: this shows
that, unlike the $W$ data discussed above, the LHC jet data included in
the NNPDF2.3 fit have a very moderate impact.

We see from Figure~\ref{fig:ttbar&ggH}(b) that, even
after accounting for all uncertainties, the ABM11 PDF set is strongly
disfavored by both the LHC top-pair cross section and the world
average value of $\alpha_S(M_Z^2)$.  The ABM11 prediction of
$\sigma_{t\bar{t}}=123.3\pm5.5(\mathrm{PDF+}\alpha_S)^{+6.2}_{-5.6}
(\textrm{scales})^{+3.6}_{-3.5}(m_t)~\textrm{pb}$
is almost 30\% below the measured cross section of
$\sigma_{t\bar{t}}=173.3\pm 10.1~\textrm{pb}$, whereas adding all
experimental and theoretical uncertainties in quadrature still gives a
total uncertainty of less than 8\%.  We note also from
Figure~\ref{fig:ttbar&ggH} that the HERAPDF1.5 NNLO predictions have a
very large model uncertainty in the upwards direction, due to varying the
minimum $Q^2$ cut from the default value of $Q^2_{\rm min} = 3.5~{\rm
GeV}^2$ to a slightly higher value of $Q^2_{\rm min} = 5~{\rm
GeV}^2$.  This sensitivity is not observed in global fits,
where the Tevatron jet data stabilize the fit and so lessen
sensitivity to the fine details of the treatment of the DIS
data~\cite{Thorne:2011kq,Watt:2012np}.

\subsubsection{Combination of Results from Different PDF Groups}
\label{sec:combination}

Whereas it is often advisable to check experimental results against
predictions obtained using the widest available set of PDFs, there are
situations where a unique reliable prediction is needed. Typical
examples are searches for new physics, or acceptance calculations,
where one does not want to inflate uncertainties unnaturally, but also
does not want to mistake an underestimated systematic effect for a new
physics effect, as has sometimes happened in the past.

This suggests that use of PDF sets based on significantly smaller
data sets is not
advisable, as these necessarily have larger uncertainties (and indeed
sometimes have appeared as outliers, as seen above). As mentioned in
Section~\ref{sec:intro}, the PDF4LHC group recommended
therefore~\cite{Botje:2011sn} the use of the CTEQ, MSTW and NNPDF
PDFs based on global fits for these purposes at the LHC. This
recommendation was adopted by the Higgs working
group~\cite{Dittmaier:2011ti}, and used for Higgs searches and
discovery~\cite{Aad:2012gk,Chatrchyan:2012gu}.

\begin{figure}
  (a)\hspace{0.5\textwidth}(b)\hfill\\
  \centerline{\includegraphics[width=0.5\textwidth]{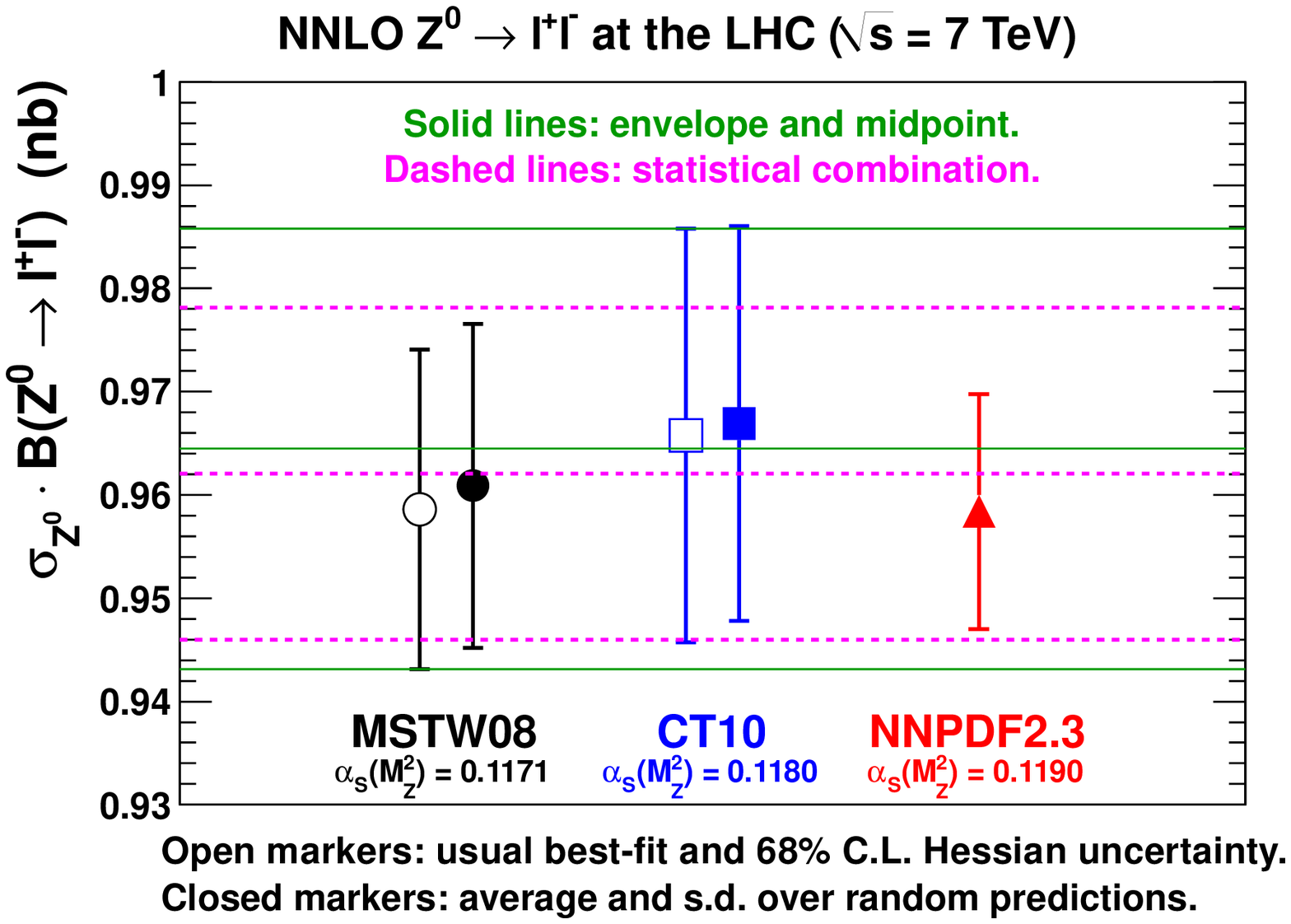}%
    \includegraphics[width=0.5\textwidth]{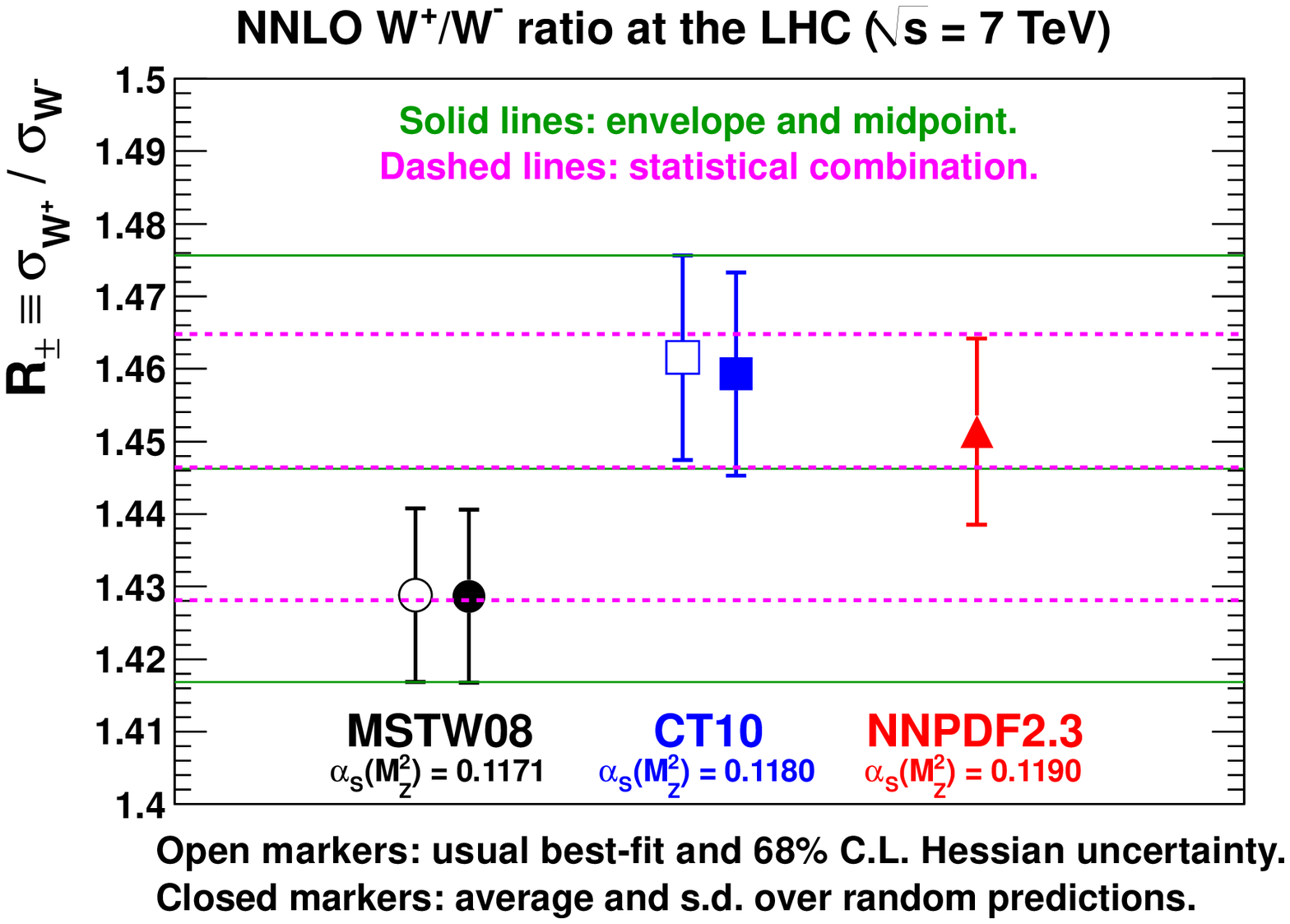}}
  (c)\hspace{0.5\textwidth}(d)\hfill\\
  \centerline{\includegraphics[width=0.5\textwidth]{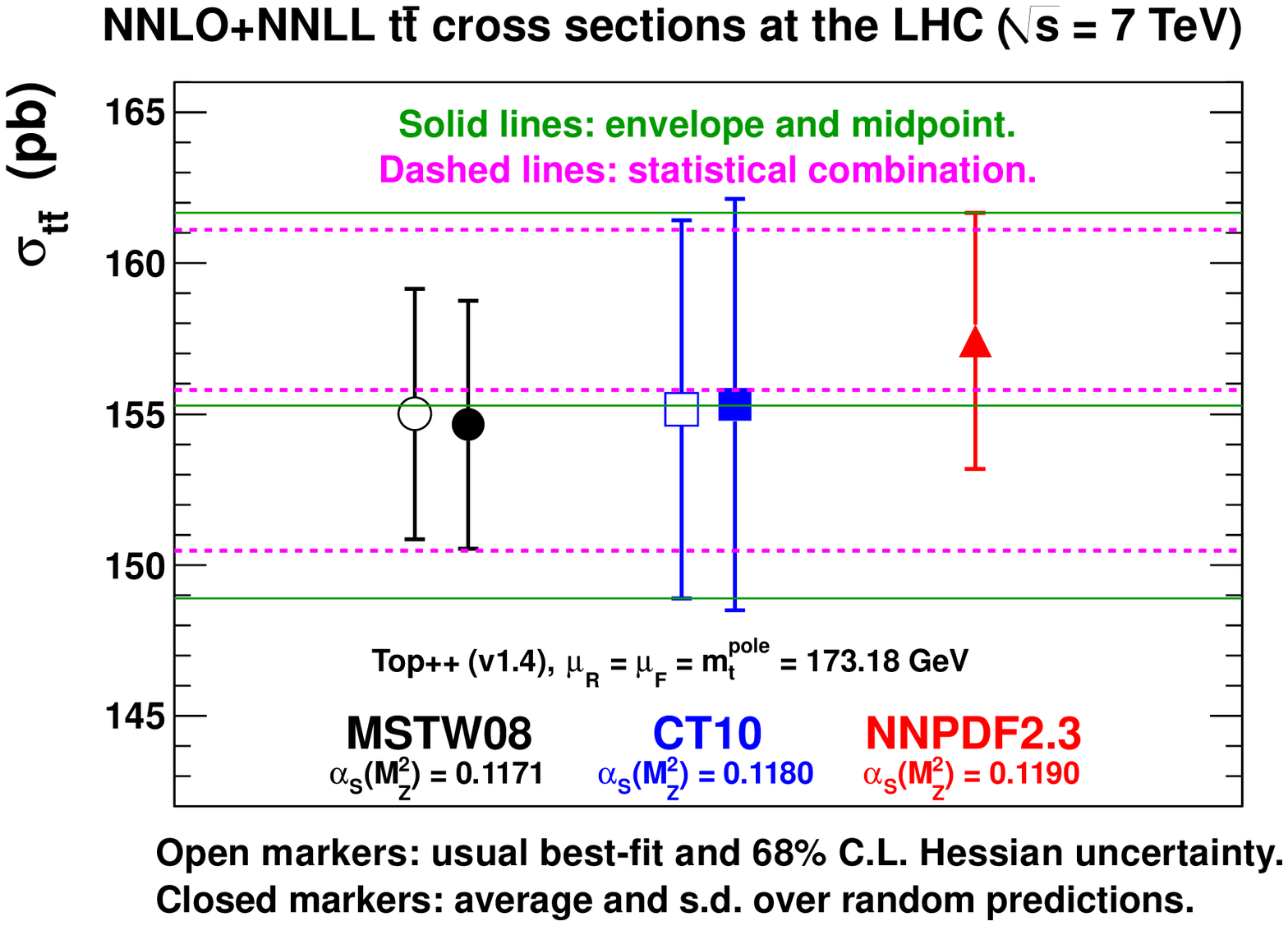}%
    \includegraphics[width=0.5\textwidth]{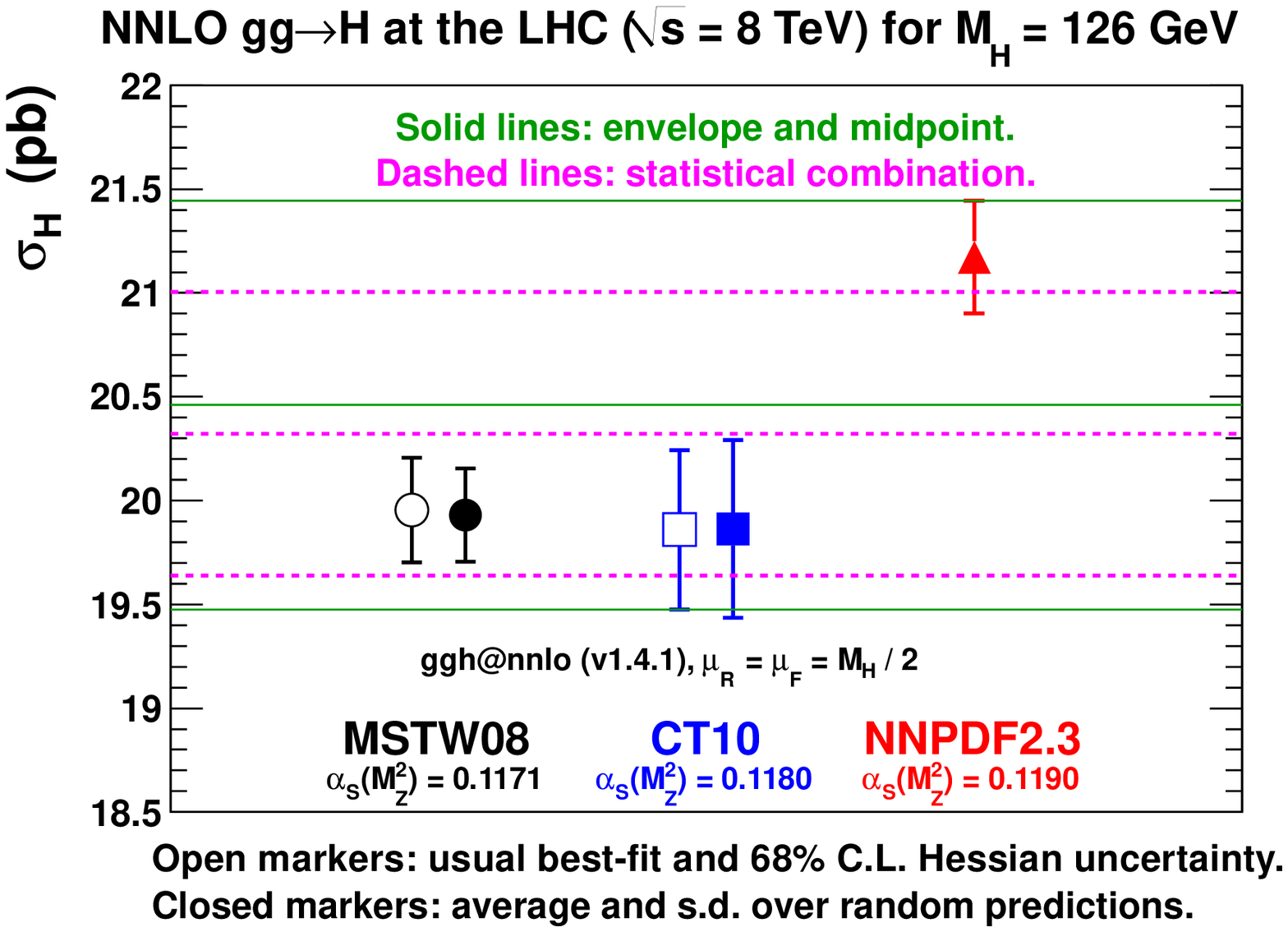}}
  \caption{NNLO (a)~$Z^0$, (b)~$W^+/W^-$, (c)~$t\bar{t}$ and
    (d)~$gg\to H$ cross sections from MSTW08, CT10 and NNPDF2.3,
    combined either by taking the envelope of the three predictions, 
    or from the statistical combination of 100 random predictions
    from each group.}
  \label{fig:combination}
\end{figure}

The problem however arises of the best way to combine results from
different groups. The PDF4LHC group originally recommended taking an envelope
of various predictions, i.e., taking as a 68\% C.L.~the range between
the top of the highest one-sigma bands, and the bottom of the lowest,
with the midpoint as a  central value. This choice is simple to
implement, but it has no clear-cut statistical meaning. A better
option is to take a statistical combination as originally proposed in
Reference~\cite{Forte:2010dt} and discussed in
Section~\ref{sec:pdfunc}.  Results obtained either way are shown in
Figure~\ref{fig:combination}, where in order to construct the
statistical combination 100 Monte Carlo replicas were generated for
MSTW08 and CT10 from the original Hessian sets using the method described in
Reference~\cite{Watt:2012tq}; see Equation~\ref{eq:randpdf}. It is
clear that the two methods
actually produce fairly similar results, though the envelope method is
somewhat more conservative, especially when there is some disagreement
between predictions.

A separate issue is how to treat the $\alpha_S$ uncertainty. The
original PDF4LHC prescription recommended a very conservative approach
in which one takes the envelope of three PDF+$\alpha_S$
uncertainties, each centered at a different central value of
$\alpha_S$: this was motivated by the feeling that the PDG $\alpha_S(M_Z^2)$
uncertainty of $\Delta\alpha_S=0.0007$ might be somewhat underestimated.
The uncertainties shown in Figure~\ref{fig:combination}
are PDF-only uncertainties for a
fixed $\alpha_S$ value.  However, results for the three PDF sets are
obtained using the three different central values shown in
Figure~\ref{fig:alphaSMZ}, so the combination implicitly
includes an additional uncertainty arising from the different
$\alpha_S$ values of $\alpha_S(M_Z^2)=0.118\pm0.001$, without going
into the complication of computing explicit PDF+$\alpha_S$
uncertainties on each separate prediction.

\subsection{PDF Constraints From the LHC}
\label{sec:constraints}

The LHC data have already started bringing in new information on
PDFs. This is expected to be even more the case in the coming years as
more processes will be studied thoroughly. Indeed, as mentioned in
Section~\ref{sec:chisq}, full information on the correlations between
the systematic uncertainties is necessary if data are to be
used for PDF determination. This is at present only available for a
handful of measurements (in particular
those used in NNPDF2.3, see Figure~\ref{fig:nnpdf23kin}), but this
situation is rapidly changing.
In fact, it is likely that in the next one or two decades most
information on PDFs, which will be crucial for new physics searches at
the LHC, will be coming from the LHC itself. The only possible
exception is if the proposed Large Hadron--electron Collider
(LHeC)~\cite{AbelleiraFernandez:2012cc}
were to be built. This would lead to a very substantial increase in
knowledge of high-energy deep-inelastic scattering which, using
separate information on charged- and neutral-current processes,
together with charm tagging, would
allow complete flavor separation from DIS alone (as discussed in
Section~\ref{sec:data}), while the large lever arm in $Q^2$ would
allow an accurate determination of the gluon from scaling
violations. More accurate results on flavor separation could only be
obtained at a neutrino factory~\cite{Mangano:2001mj}.

\subsubsection{Light Flavors}

The strongest constraint on light flavor PDF at the LHC comes from the
combination of rapidity distributions for the production of various
gauge bosons: as repeatedly mentioned, a rapidity distribution fixes
entirely the leading-order parton kinematics.  If full information
on the correlation between different processes is retained, a global
fit including all of them effectively uses the information provided by
all the various cross-section ratios that are sensitive to different
PDF combinations, such as Equations~\ref{eq:WoverZ} and \ref{WpWm}, or,
equivalently the asymmetry
\begin{equation}
 \frac{
   \sigma_{W^+}-\sigma_{W^-}}{\sigma_{W^+}+\sigma_{W^-}}\sim\frac{u_v(x_1)
   -d_v(x_1)}{u(x_1)+d(x_1)}.
\end{equation}
It is important to observe that many sources of systematics are
common (such as for instance the normalization) to these
cross sections and cancel in the ratio: hence the availability of full
correlations leads to potentially much more precise results. Given
that the LHC energy is being increased in stages, it is also possible
to form ratios or double ratios between measurements at different
energies which further increase the potential for
precision~\cite{Mangano:2012mh}. 

The most discriminating data on $W$ production so far are the CMS
asymmetry data~\cite{Chatrchyan:2012xt}, included in the NNPDF2.3 fit, and
which are mostly responsible for the sizable reduction in uncertainty
seen in Figure~\ref{fig:wz}(b) when comparing the NNPDF2.3 and
NNPDF2.3noLHC fits. These measurements have in particular shown that
the asymmetry is underestimated by the MSTW08 fit,
implying that $u_v-d_v$ is too small
at $x\sim M_W/\sqrt{s}\sim 0.01$.
\begin{figure}
  (a)\\
  \centerline{\includegraphics[width=0.94\textwidth]{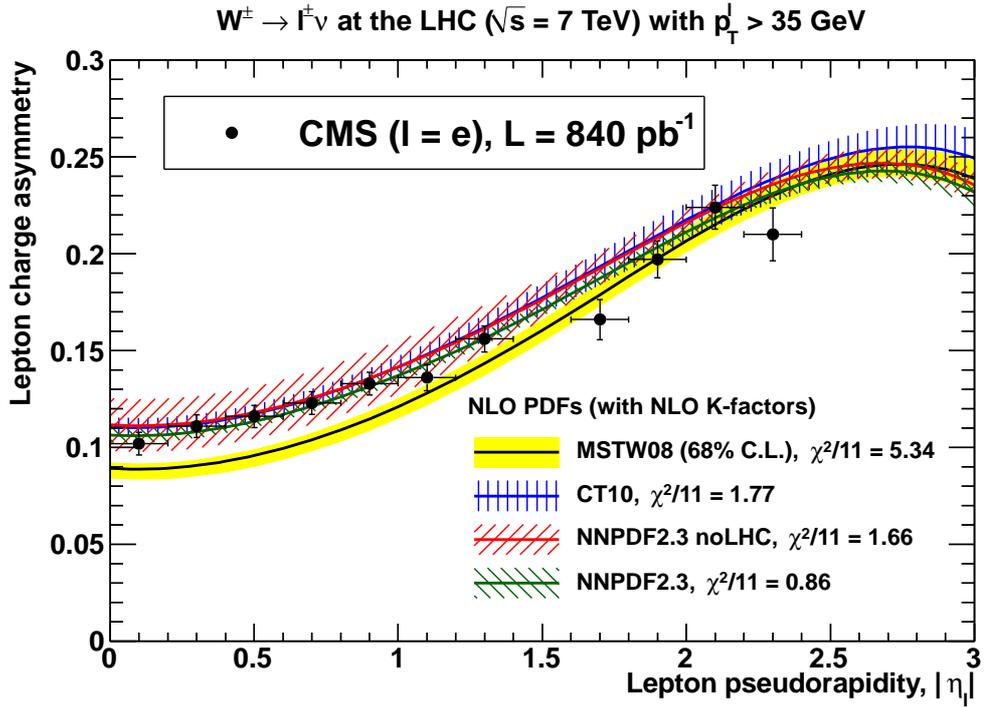}}
  (b)\\
  \centerline{\includegraphics[width=0.94\textwidth]{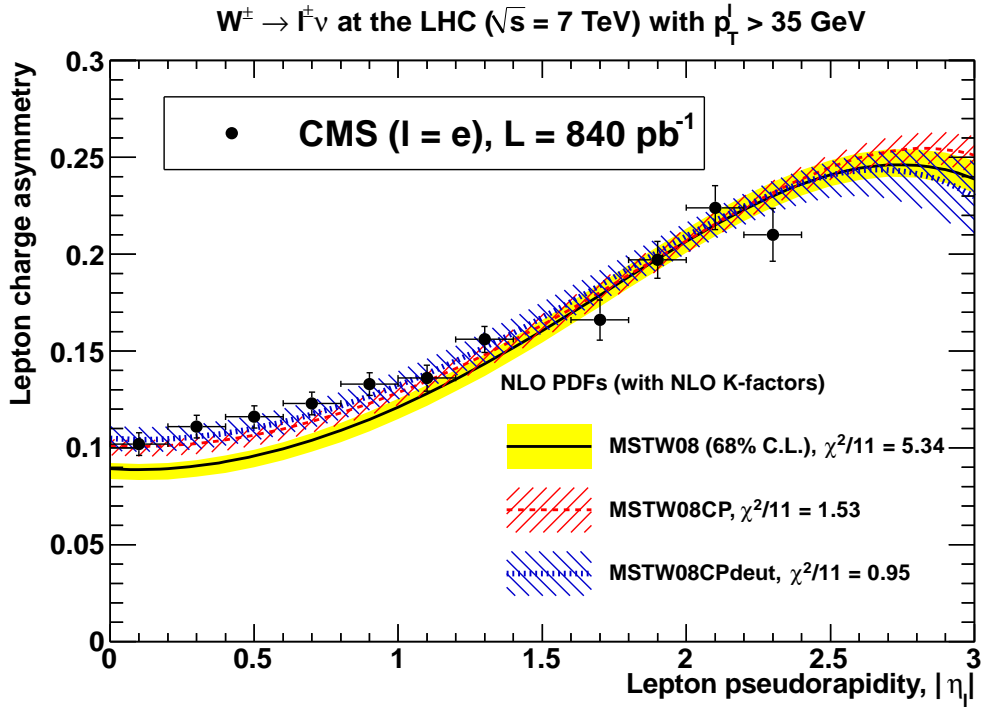}}
  \caption{Description of CMS electron asymmetry~\cite{Chatrchyan:2012xt}
    using (a)~various NLO PDF sets, and (b)~improved variants of the
    MSTW08 analysis~\cite{Martin:2012da}.}
  \label{fig:cms35}
\end{figure}
In Figure~\ref{fig:cms35} we show the CMS data from
Reference~\cite{Chatrchyan:2012xt} with
$p_T^\ell>35$~GeV. It is apparent that
inclusion of these data in the NNPDF2.3 analysis resulted
in a more than three-fold
reduction in the NLO PDF uncertainty on the asymmetry at central pseudorapidity.
They prompted a reexamination of the
PDF parametrization (and also deuteron corrections) used in the MSTW08 fit,
where an extended Chebyshev parametrization form and more flexible
deuteron corrections were found to automatically improve the
description of the CMS electron asymmetry data~\cite{Martin:2012da};
see Figure~\ref{fig:cms35}(b).  (Figure~\ref{fig:cms35}
shows only NLO PDF sets since the main studies of
Reference~\cite{Martin:2012da} were done at this order).  
This is an example of how present and future LHC data may help in
resolving discrepancies which are still present between PDF sets.

In addition to $W$ and $Z$ production, measurements of Drell--Yan processes 
($Z/\gamma^*\to \ell\ell$ or $W\to\ell\nu$) away from the resonance peaks
at $M_{\ell\ell}=M_Z$ or $M_{\ell\nu}=M_W$ in either direction can provide
complementary constraints on different PDF combinations and $x$ values.

\subsubsection{Gluon}

The NNPDF2.3 analysis also includes inclusive jet
data from ATLAS~\cite{Aad:2011fc}.  These could in principle
result in a reduction of uncertainties on the gluon distribution at
the $x$ values in the intermediate region shown in
Figure~\ref{fig:nnpdf23kin}, though in practice their impact is very
moderate, as seen from the luminosity plot in Figure~\ref{fig:gglumi}
and the Higgs and top production cross sections in Figure~\ref{fig:ttbar&ggH}.
This situation is likely to change in the future, as jet data become more
abundant and precise. For example, the recent CMS jet
data~\cite{Chatrchyan:2012bz} are expected to place more stringent
constraints on the gluon distribution
than the available ATLAS data~\cite{Aad:2011fc}. 

As is clear from Figure~\ref{fig:ttbar&ggH},
data for top-pair production, especially
differential distributions~\cite{Chatrchyan:2012qka}, will provide
stringent constraints on the gluon. The process is only known at NLO at
the differential level, but at the inclusive level the full NNLO
result is just around the corner~\cite{Czakon:2012pz}.
Eventually, once all aspects of the observed Higgs-like boson are understood,
even Higgs production itself (which at least in the Standard Model is known
up to NNLO) could be perhaps the most sensitive probe of the gluon
distribution.

Other LHC data may be used to constrain the gluon distribution. A
particularly clean probe might come from gauge-boson $p_T$
distributions, because the final-state $W$ or $Z$ can acquire a $p_T$
dependence only if at least one parton is radiated. The values of $p_T$
accessible at the LHC are large enough that it should be possible to
obtain significant constraints even by imposing a cut in order to
avoid the low $p_T$ region where QCD resummation is mandatory and
non-perturbative effects might become relevant. The impact on PDFs of these
data, some of which are already
available~\cite{Aad:2011gj,Chatrchyan:2011wt}, has not been studied
yet.  However, as with the closely related observable of $W,Z$+jet production,
only a NLO ($\mathcal{O}(\alpha_S^2)$) calculation is available.

Another classic process which may be used to constrain the gluon
distribution is prompt photon production, and the related photon+jet
production process, to which gluons contribute at leading order. 
Inclusion of present-day LHC prompt photon data
would only lead to a moderate reduction of order of $\sim20\%$ of the
gluon uncertainty~\cite{d'Enterria:2012yj}  at medium-small
$x\sim0.02$, with photon+jet data having an even milder
impact~\cite{Carminati:2012mm}, but future data are likely to be much
more  constraining. Here, the main bottleneck is that theoretical
predictions are only available up to NLO, and higher order corrections
are likely to be large.
Diphoton production is known up to NNLO~\cite{Catani:2011qz}, but it
is less constraining in that the gluon contribution only starts at NLO.

A potential option to get a handle on the gluon is the study of $W$
polarization~\cite{Chatrchyan:2011ig,ATLAS:2012au,Stirling:2012zt}, which
is similar to the $W$ $p_T$ distribution discussed above, but subject to
small QCD corrections~\cite{Bern:2011ie}.

\subsubsection{Strangeness and Heavy Quarks}

As already discussed in Section~\ref{strangesec}, $W$ production also
provides a handle on strangeness, and in fact Tevatron $W$ production
data were found~\cite{Ball:2010de} to have a significant impact on the
$s-\bar s$ distribution. It has been argued~\cite{Aad:2012sb} that a
fit only to inclusive $W^\pm$ and $Z$ differential cross
sections~\cite{Aad:2011dm}, combined with inclusive DIS data from
HERA,  can significantly constrain the strange content of the proton,
finding apparently no strange suppression, contrary to previous
determinations from CCFR/NuTeV dimuon cross sections ($\nu N\to \mu\mu
X$).  However, these conclusions are weakened in a similar NNPDF study
using a more flexible parametrization form~\cite{Ball:2012cx}, or
alternatively by incorporating a suitable
tolerance~\cite{Watt:2012tq}, and it seems that present-day LHC $W$
production data only have a very minor impact on strangeness, but this
is again likely to change in the near future.

A much more direct handle on strangeness  at the LHC comes from $W$
production with an associated charm-tagged jet, where the dominant
partonic subprocesses are $\bar{s}\,g\to W^+\,\bar{c}$ and $s\,g\to
W^-\,c$.  A first preliminary measurement has been made by
CMS~\cite{CMS:2011lqa} of the cross-section ratios
$R_c\equiv\sigma(W+c)/\sigma(W+{\rm jets})$, probing the strange
content of the proton relative to other light-quark flavors, and
$R_c^\pm\equiv\sigma(W^++\bar{c})/\sigma(W^-+c)$, potentially probing
the strange asymmetry.  With more precise measurements to come,
including differential distributions, the $W$+charm process should
enable powerful constraints to be made on the $s$ and $\bar{s}$
distributions~\cite{Stirling:2012vh}.   The main limitation here is
that only NLO results are available for this process.

Similarly,  $Z$ production in association with a tagged charm jet
will provide  significant constraints on the charm
distribution~\cite{Stirling:2012vh}, and likewise, $Z$ with a $b$ jet
on the $b$ distribution.

%%%%%%%%%%

\section{THE FUTURE OF PDF DETERMINATION}
\label{sec:summary}

Parton distributions have become increasingly relevant with
the advent of the LHC.
After the landmark discovery of a Higgs-like
boson in July 2012~\cite{Aad:2012gk,Chatrchyan:2012gu}, focus now shifts to
characterizing the properties of this new particle, as well as increasingly
difficult searches for indications of other new
physics. As a consequence, demand will grow to reduce the unavoidable
uncertainties associated with the PDFs in calculations of both
signal and background.
On the one hand, it will be necessary to bring under complete control 
the uncertainties in the region of electroweak symmetry breaking,
i.e., the region
of $x$ and $Q^2$ which is probed by Higgs production.  In this region, as
mentioned in Section~\ref{sec:comp}, uncertainties are in principle small,
but in practice for some PDFs such as the gluon and even more the strange,
discrepancies between different sets at the one or two sigma level are
seen and not fully understood. On the other hand, searches for new physics
will involve heavy final states, and thus, because of Equation~\ref{taudef},
they will involve knowledge of PDFs in the large $x\gtrsim0.5$ region
where they are currently very poorly known.
To achieve these goals, it will be necessary to construct PDFs which satisfy
a number of criteria, that not so long ago~\cite{Forte:2010dt} characterized
an ideal PDF determination, but in the LHC era have become necessary
requirements, namely, in decreasing order of importance:

\begin{enumerate}

\item The range and precision of \textbf{data sets} must be as wide as
  possible, cover currently unexplored kinematic regions, and include
  new LHC processes, such as those discussed in Section~\ref{sec:constraints},
  which will gradually remove current discrepancies between PDF determinations.

\item The \textbf{parametrization} should be sufficiently general and
  demonstrably unbiased, either by using a sufficiently large number of
  parameters, or by careful a posteriori checks of parametrization independence.

\item The \textbf{experimental uncertainties} should be understood and carefully
  propagated, and in particular the statistical meaning of the procedures
  that are being adopted to determine the PDF uncertainties should be
  understood: specifically, the choice of tolerance, and the determination of
  the optimal fit when using a very flexible parametrization.

\item Computations should be performed at the highest available
  \textbf{perturbative order}, and in particular, at the order which
  is subsequently to be used in the computation of partonic cross sections.
  This is currently NNLO, but the need for the inclusion of various kinds
  of all-orders resummation is becoming increasingly important. For example,
  computations which include Sudakov resummation are already being used in the
  computation of the Higgs production cross section~\cite{deFlorian:2012yg},
  and for consistency resummed computations should also be used in PDF
  determination~\cite{Corcella:2005us}: this becomes especially important
  in the large $x\gtrsim0.5$ region that will be probed in searches for
  new physics. Also, the inclusion of resummation effects will be increasingly
  important in the construction of PDFs to be used in fixed-order
  calculations matched to parton showers in Monte Carlo event generators.
  Finally, PDFs including electroweak corrections will have to be
  constructed~\cite{Martin:2004dh}.

\item The treatment of \textbf{heavy quarks} will have to include
  mass-suppressed terms in the coefficient functions, while also
  resumming logarithmically enhanced terms via the evolution equations.
  Such a treatment, like the schemes discussed in Section~\ref{sec:pert},
  is a minimum requirement: this is currently standard for DIS, but
  applications to hadronic observables are so far limited.  Also, the
  dependence of results on the choice of value for the heavy-quark masses
  will have to be studied more systematically, possibly using the
  perturbatively more stable $\overline{\rm MS}$ mass
  definition~\cite{Alekhin:2010sv}, with PDF sets made available
  for several values of the heavy-quark masses.

\item The \textbf{strong coupling} $\alpha_S$, in addition to being determined
  simultaneously with PDFs, should also be decoupled from the PDF
  determination, with PDF sets available for a range of fixed
  $\alpha_S$ values, and full PDF uncertainty determination for each
  value of $\alpha_S$.

\item Estimate of \textbf{theoretical uncertainties} 
  will have to be performed together
  with PDF sets, and such uncertainties will have to be provided each time
  they become comparable with other sources of PDF uncertainty. This is
  presently an almost unexplored territory.

\end{enumerate}

We predict that, as the needs of precision physics at the LHC develop,
more and more of these features will become accepted standards.

\acknowledgments

We are indebted to all participants of the PDF4LHC workshops, in particular
A.~de~Roeck, A.~Glazov, J.~Huston, P.~Nadolsky, J.~Pumplin,
and to all the members of the MSTW and NNPDF collaborations,
especially J.~Rojo (whom we also thank for a critical reading of
the manuscript) and R.~Thorne for innumerable discussions on the subject of
this review.

%%%%%%%%%%

\bibliographystyle{JHEP}

\bibliography{ForteWattArxiv}

\end{document}